\documentclass[usenatbib,useAMS]{mn2e}
\usepackage{times}
\usepackage{epsfig}

\renewcommand{\vec}[1]{\bmath{#1}}
\def\earth{\oplus}

\begin{document}
\title{Revealing stellar brightness profiles by means of microlensing fold caustics}
\author[M. Dominik]{M. Dominik \\
University of St Andrews, School of Physics \& Astronomy, North Haugh,
St Andrews, KY16 9SS}

\maketitle

\begin{abstract}
With a handful of measurements of limb-darkening coefficients, galactic
microlensing has already proven to be a powerful technique for studying
atmospheres of distant stars.
Survey campaigns such as OGLE-III are capable of providing
 $\sim\,10$ suitable target stars per
year that undergo microlensing events involving passages over
the caustic created by a binary lens, which last from a few hours to
a few days and allow to resolve the stellar atmosphere by frequent broadband photometry.
For a caustic exit lasting 12~h and a photometric precision of 1.5~\%, 
a moderate sampling interval of 30~min (corresponding to $\sim\,$25--30 data points) is
sufficient for providing a reliable measurement of the linear limb-darkening
coefficient $\Gamma$ with an uncertainty of 
$\sim 8\,$\%, which reduces to  $\sim 3\,$\% for a reduced sampling interval of
6~min for the surroundings of the end of the caustic exit.
While some additional points over the remaining parts of the lightcurve are highly
valuable, a denser sampling in these regions provides little improvement.
Unless an accuracy of less than $5\,$\%
is desired, limb-darkening coefficients for several filters can be obtained
or observing time can be spent on other targets during the same night.
The adoption of an inappropriate stellar brightness profile as well
as the effect of acceleration between source and caustic yield 
distinguishable characteristic systematics
in the model residuals.
Acceleration effects are unlikely to affect the
lightcurve significantly for most events, although a free acceleration parameter
blurs the limb-darkening measurement if the passage duration cannot be accurately determined.
\end{abstract}


\section{Introduction}

The light we receive from a star originates from layers at different depths 
from its surface, 
where the contribution of inner layers decreases from the center
towards the limb yielding
the phenomenon of limb darkening of the observed surface brightness across the
stellar face, which reflects
the temperature of the 
stellar atmosphere as function of distance from the center.
Decreasing temperatures from the center to the surface cause
a stronger effect for shorter wavelengths, making stars bluer near their center
and redder near their limb.

Using caustics of gravitational lenses to measure the brightness profile
of closely-aligned background sources was proposed by \citet{SW1987}
with the view toward active galactic nuclei, 
and several authors have later discussed its
application to source stars undergoing galactic microlensing events that
involve fold-caustic passages \citep[e.g.][]{Rhie:LD,GG:detLD}.
Making use of both
the large total magnification and a strong differential
magnification across the
stellar face during the caustic passage, galactic microlensing has
successfully been demonstrated
to be a powerful
technique for revealing the brightness profile of distant stars by 
providing several measurements of limb-darkening coefficients characterizing
linear or square-root laws \citep{PLANET:M28, PLANET:M41, PLANET:O23,
joint, PLANET:EB5, Yooetal} with uncertaintes of a few percent.

As discussed in detail by \citet{Do:Fold}, the lightcurve of a source star
in the vicinity of a fold caustic is completely characterized by local
properties, so that its brightness profile can be obtained already from
data taken around the caustic passage 
without the need to find a model for the full lightcurve, which describes
all properties of the lens system and involves a larger number of parameters along with 
possible parameter ambiguities. Nevertheless, data outside the caustic passage 
region provide stronger constraints on the differential blending between
observing sites and filters as well as a better assessment of acceleration
effects between source and caustic, which can be caused by the parallactic
motion of the Earth around the Sun or by orbital motion of the binary lens
or a possible binary source. This additional information can be used to
reduce uncertainties in the measurement of the stellar brightness profile
and its characterizing limb-darkening coefficients.  

Some aspects of the measurement of limb-darkening coefficients from
microlensing events involving fold-caustic passages have previously been
studied by \citet{Rhie:LD}. This article provides a deeper and 
furthergoing discussion on this issue and on some general properties of 
lightcurves of microlensed sources near fold caustics with respect to revealing
their brightness profiles. 

\begin{figure*}
\includegraphics[width=168mm]{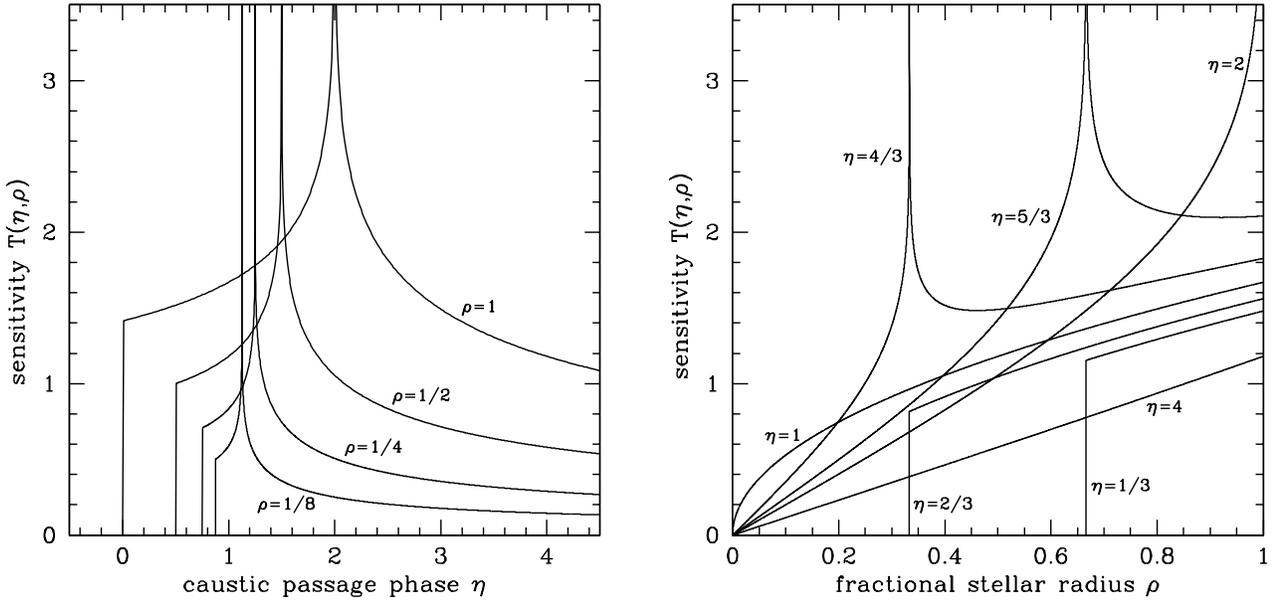}
\caption{The sensitity ${\bmath {\mathcal T}}(\eta, \rho)$ of the 
fold caustic to the shape of the stellar brightness profile $\xi$
as a function of the caustic passage phase 
$\eta = \pm\,(t-t_\rmn{f}^\star)/t_\star^\perp$ (left) and the fractional stellar radius
$\rho$ (right), as given by Eq.~(\ref{eq:sensfunc}).}
\label{fig:sensitivity}
\end{figure*}

Sect.~\ref{sec:lightcurve} addresses the relation between the lightcurve and
the stellar brightness profile and discusses the
sensitivity of microlensing  
as function of phase during the caustic passage and fractional radius
of the source star.
With linear limb darkening as example,
Sect.~\ref{sec:linear} deals with parametrized profiles.
A simulation of typical sampled data for
different sampling rates is described in Sect.~\ref{sec:sampled}, which are
used in Sect.~\ref{sec:reconstruction} to investigate
the amount of information about the
linear limb-darkening coefficient contained in
different regions of the lightcurve.
In Sect.~\ref{sec:wrongprofile}, consequences arising from
the adoption of an inadequate limb-darkening profile profile are discussed,
whereas Sect.~\ref{sec:acceleration} addresses the influence
of an effective acceleration between source
and caustic.
The paper finishes with a summary of the results and recommendations on the 
observing strategy
in Sect.~\ref{sec:summary}.

\section{Intensity profile and lightcurve}
\label{sec:lightcurve}

Let us consider a gravitational lens of total mass $M$ at distance $D_\rmn{L}$ from
the observer acting on a source at distance $D_\rmn{S}$. A characteristic scale 
is then given by the angular Einstein radius
\begin{equation}
\theta_\rmn{E} = \sqrt{\frac{4GM}{c^2}\,\frac{D_\rmn{S}-D_\rmn{L}}{D_\rmn{L}\,D_\rmn{S}}}\,,
\end{equation}
while its corresponding physical radius at the lens distance is the Einstein radius 
$r_\rmn{E} = D_\rmn{L}\,\theta_\rmn{E}$.

With the combined magnification due to the two critical images of
a point-like source at a distance $u_\perp \theta_\rmn{E}$
from a fold caustic given by
\begin{equation}
A_\rmn{crit}^\rmn{p}(u_\perp) = \left(\frac{R_\rmn{f}}{u_\perp}
\right)^{1/2}\,
\Theta(u_\perp)\,,
\end{equation}
where $R_\rmn{f}$ describes the caustic strength, $u_\perp > 0$ ($u_\perp < 0$) for
a source inside (outside) the caustic, and $\Theta(x)$ is the step function,
the corresponding magnification for 
a circle of angular radius $r \theta_\rmn{E}$ with its center at  $u_\perp \theta_\rmn{E}$ 
reads
\begin{equation}
A_\rmn{crit}^\rmn{circle}(u_\perp, r) = 
\left(\frac{R_\rmn{f}}{r}
\right)^{1/2}\,j\left(\frac{u_\perp}{r}\right)\,,
\end{equation}
where
\begin{equation}
j(z) = \left\{\begin{array}{lcl}
0 & \mbox{for} & z \leq -1 \\
\frac{\sqrt{2}}{\upi}\,K\left(\sqrt{\frac{1+z}{2}}\right)
& \mbox{for} & -1 < z < 1 \\
\frac{2}{\upi\,\sqrt{1+z}}\,K\left(\sqrt{\frac{2}{1+z}}\right)
& \mbox{for} & z > 1
\end{array} \right.\,,
\end{equation}
and $K$ denotes the complete elliptical integral of first kind
\citep[c.f.][]{GG:detLD}.

Consider now a star with angular radius $\rho_\star \theta_\rmn{E}$ 
whose leading limb enters or whose trailing 
limb exits a fold caustic
at time $t_\rmn{f}^\star$ and which needs 
the timespan $2\,t_\star^\perp$ to cross. The parameter 
\begin{equation}
\eta = \pm\,\frac{t-t_\rmn{f}^\star}{t_\star^\perp} 
= \frac{u_\perp}{\rho_\star} + 1
\end{equation}
then describes the phase of the caustic passage, where the upper
sign corresponds to a caustic entry and the lower sign corresponds
to a caustic exit. The source
is completely outside the caustic for $\eta < 0$, completely inside
the caustic for $\eta > 2$, and passes the caustic for $0 \leq \eta \leq 2$,
where $\eta = 1$ corresponds to the center of the star crossing the
caustic. As will be shown in the next section, the reference time
$t_\rmn{f}^\star$ is preferred over the time $t_\rmn{f} = t_\rmn{f}^\star
\pm t_\star^\perp$ when the source crosses the caustic,
because it can be immediately identified from a distinctive
feature in the lightcurve.

With $\rho$ denoting the fractional stellar radius, the
stellar brightness profile for a given filter reads 
\begin{equation}
I(\rho) = \overline{I}\,
\xi(\rho)\,,
\end{equation} 
where
$\overline{I}$ denotes the average surface brightness and 
$\xi(\rho)$ is a dimensionless function describing the
radial profile, so that
\begin{equation}
\int\limits_0^1 \xi(\rho)\,\rho\,\rmn{d}\rho = \frac{1}{2}\,.
\end{equation}

The magnification of the critical images takes the form
\begin{equation}
A_\rmn{crit}(\eta;\xi) = 
\left(\frac{R_\rmn{f}}{\rho_\star}\right)^{1/2}\,
G_\rmn{f}^\star(\eta;\xi)\,,
\end{equation}
and the characteristic caustic profile function $G_\rmn{f}^\star(\eta;\xi)$ can be written
as
\begin{equation}
G_\rmn{f}^\star(\eta;\xi) = \int\limits_{0}^{1} {\bmath{\mathcal T}} (\eta,\rho)\;\xi(\rho)\, \rmn{d}\rho\,,
\label{eq:response}
\end{equation}
where ${\bmath{\mathcal T}} (\eta,\rho)$ describes the response
produced by the caustic to the brightness profile
$\xi(\rho)$. 
In general,
the determination of the brightness profile $\xi(\rho)$ 
involves an inversion of Eq.~(\ref{eq:response}), which constitutes a Fredholm integral
equation of first kind. The solution of this problem is non-trivial
and has been discussed
in some detail by \citet{Hey:Fredholm}. 

Since 
\begin{equation}
A_\rmn{crit}(\eta; \xi) = 2\,\int\limits_0^1
A_\rmn{crit}^\rmn{circle}\left(\frac{\eta-1}{\rho_\star},
\rho\,\rho_\star\right)\,\xi(\rho)\,\rho\,\rmn{d}\rho\,,
\label{eq:afromcirc}
\end{equation}
one finds that 
\begin{equation}
{\bmath{\mathcal T}} (\eta,\rho) = 2\,\rho^{1/2}\,j\left(\frac{\eta-1}{\rho}\right)\,.
\label{eq:sensfunc}
\end{equation}
Eq.~(\ref{eq:afromcirc}) reflects the fact that, in order to account for the fraction of
light provided by each circle, the surface brightness at the corresponding fractional radius has
to be weighted by its circumference,  which is proportional to $\rho$.

The function ${\bmath{\mathcal T}} (\eta,\rho)$ is shown in
Fig.~\ref{fig:sensitivity} as a function of the caustic passage phase
$\eta$ for selected 
values of the fractional stellar radius $\rho$, and as a function of 
the fractional stellar radius $\rho$ for selected values of the caustic 
passage phase $\eta$. 
As long as the circle corresponding to the fractional radius $\rho$ is
outside the caustic, only its non-critical images contribute to its
light, so that ${\bmath{\mathcal T}} (\eta,\rho)$ is zero for
for $\eta < \eta_\rho^{\rmn{out}} = 1-\rho$, where the caustic is touched and
the sensitivity jumps to
the value $(2\rho)^{1/2}$.
The sensitivity increases with $\eta$
for $\eta_\rho^{\rmn{out}} < \eta < 
\eta_\rho^{\rmn{in}}$ until the circle is completely inside the
caustic at $\eta_\rho^{\rmn{in}} = 1+\rho$, where 
${\bmath{\mathcal T}}(\eta,\rho)$ diverges.
For $\eta > \eta_\rho^{\rmn{in}}$,
the sensitivity asymptotically 
approaches zero as ${\bmath{\mathcal T}} (\eta,\rho) \simeq \rho\,\eta^{-1/2}$.

The sensitivity of the lightcurve 
to the brightness profile at different fractional radii
$\rho$ for a given caustic passage phase $\eta$ qualitatively depends on whether
the souce is completely outside the caustic ($\eta < 0$), completely 
inside the caustic ($\eta = 2$), or in the process of passing it
with its center outside ($0 \leq \eta < 1$), inside ($1 < \eta \leq 2$), or
at the caustic ($\eta = 1$). For any caustic passage phase $\eta$, the sensitivity
${\bmath{\mathcal T}}(\eta,\rho)$
vanishes at the source center ($\rho = 0$), while it vanishes identically for any fractional
radius $\rho$ if the source is completely outside the caustic ($\eta < 0$), where
none of the corresponding circles have critical
images. With parts of the source 
moving inside
while the source center
is still outside the caustic ($\eta < 1$), 
only the outer parts of the source with
$\rho > \rho_\eta^\rmn{out} = 
1-\eta$ have non-vanishing sensitivities. ${\bmath{\mathcal T}} (\eta,\rho)$ jumps
from zero to $[2(1-\eta)]^{1/2}$ at the fractional radius 
$\rho_\eta^\rmn{out}$ and increases towards larger $\rho$.
For the source center being at the caustic ($\eta = 1$), one finds that
${\bmath{\mathcal T}} (1,\rho) \propto \rho^{1/2}$. 
If the source moves further inside ($1 < \eta \leq 2$),
increasing fractional
radii subsequently dominate the sensitivity function
through their infinite peak at 
$\rho_\eta^\rmn{in} = \eta-1$, to which ${\bmath{\mathcal T}} (\eta,\rho)$ 
rises monotonically for $\rho < \rho_\eta^\rmn{in}$. For 
$\rho > \rho_\eta^\rmn{in}$, ${\bmath{\mathcal T}} (\eta,\rho)$ shows a monotonic decrease 
from the infinite peak larger $\eta$, while a local minimum exists
for smaller $\eta$, so that a decrease is followed by an increase
towards $\rho = 1$. Finally, 
for sources completely inside the caustic ($\eta > 2$), 
${\bmath{\mathcal T}} (\eta,\rho)$ rises monotonically with the fractional radius
$\rho$ from zero at the source center ($\rho = 0$) to a maximum at the limb ($\rho = 1$)
and becomes asymptotically proportional to $\rho$ for
large $\eta$, for which ${\bmath{\mathcal T}} (\eta,\rho) \simeq
\rho\,\eta^{-1/2}$.

The properties of the brightness profile sensitivity function 
${\bmath{\mathcal T}} (\eta,\rho)$ show that
photometric observations during the passage of a source star over
a fold caustic provide a one-dimensional scan of its brightness profile 
$\xi(\rho)$, 
during which each fractional radius $\rho$ is most efficiently probed
in different ways on the two occasions where its associated circle touches the
caustic, namely at $\eta_\rho^\rmn{out} = 1-\rho$, where the 
profile sensitivity function jumps from zero to the finite value 
$(2\rho)^{1/2}$, and at $\eta_\rho^\rmn{in} = 1+\rho$, where it 
becomes infinite. Therefore, outer parts of the source
are most efficiently probed near the beginning or the end of
the caustic passage, whereas inner parts are most efficiently 
probed while the center of the source passes the caustic.
Influenced by the fact that circles corresponding to
larger fractional radii have a larger circumference and therefore 
contribute more light for the same surface brightness, the integrated sensitivity of the
caustic to the brightness profile over the full course of the caustic passage
increases with fractional radius, so that data of constant quality would yield more information
about the outer than about the inner parts of the source.\footnote{For constant exposure time
during the caustic passage, larger magnifications however account for
a higher photometric precision.}

Rather than $G_\rmn{f}^\star(\eta; \xi)$ 
itself, 
one observes the lightcurve \citep[e.g.][]{Do:Fold}
\begin{eqnarray}
& & \hspace*{-1em}
m_\rmn{fold}(t)  =  m_\rmn{f}^{\star} \,- \nonumber \\
& & 2.5\,\lg\Bigg\{
1 + \frac{1}{g_\rmn{f}^\star}
\left[\left(\frac{\hat t}{t_\star^\perp}\right)^{1/2}\,
{G}^\star_\rmn{f}
\left(\pm\,\frac{t-t^\star_\rmn{f}}{t_\star^\perp};
\xi\right)\right.\,+ \nonumber \\
& & \quad +\,
{\hat G}_\rmn{other}^\star
\left(\pm
{\hat \omega}^\star_\rmn{f}\,(t-t^\star_\mathrm{f}),
{g_\rmn{f}^\star}
\right)\Bigg]
\Bigg\}
\end{eqnarray}
in the vicinity of the caustic passage, 
with $t_\rmn{f}^\star$ and $t_\star^\perp$ as defined above,
$\hat t$ being an arbitrarily chosen unit time, and
$m_\mathrm{f}^{\star}$
the observed magnitude at time $t_\rmn{f}^\star$.
With $m_\rmn{S} \leq m_\mathrm{f}^{\star}$ denoting the intrinsic source magnitude,
the parameter $g_\rmn{f}^\star$ reads
\begin{equation}
g_\rmn{f}^\star = \left(\frac{\rho_\star}{R_\rmn{f}}\right)^{1/2}
\left(\frac{\hat t}{t_\star^\perp}\right)^{1/2}\,
10^{0.4(m_\rmn{S} -m_\mathrm{f}^{\star})}\,.
\end{equation}
Finally, ${\hat G}_\rmn{other}^\star$ describes the temporal variation of the magnification 
due to
images of the source star that do not become critical if it hits the caustic, which is
measured by ${\hat \omega}^\star_\rmn{f}$ and can be approximated as
\begin{eqnarray}
{\hat G}_\rmn{other}^\star (\Lambda, {\hat g}) & = &
{\hat g}
\,\left(
\exp\left\{\Lambda/{\hat g}
\right\}-1
\right)
 \nonumber \\
& \simeq & \Lambda \quad (|\Lambda/{\hat g}| \ll 1)\,.
\end{eqnarray}

The establishment of $G_\rmn{f}^\star(\eta; \xi)$, from which the
stellar brightness profile $\xi(\rho)$ is to be inferred, therefore requires a
determination of the 5 parameters
$t_\rmn{f}^\star$, $t_\star^\perp$, 
$m_\rmn{f}^{\star}$, 
$g_\rmn{f}^\star$, and ${\hat \omega}_\rmn{f}^\star$
from a fit to the collected data, 
where in particular, $t_\rmn{f}^\star$ and
$t_\star^\perp$
determine the relation between the caustic passage phase $\eta$ and
the elapsed time $t$.

\section{Parametrized profiles}
\label{sec:linear}

\begin{figure}
\includegraphics[width=84mm]{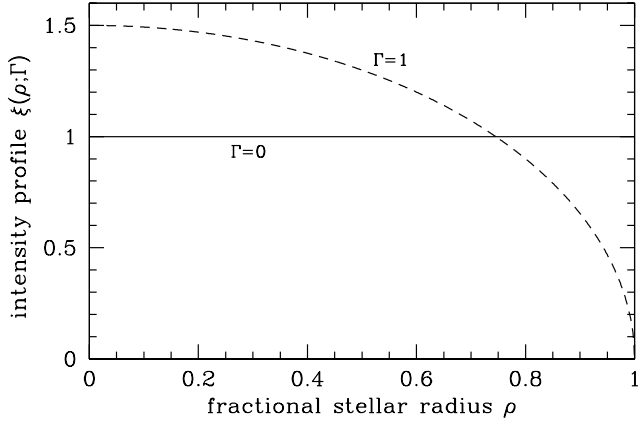}
\caption{Stellar brightness profiles $\xi(\rho; \Gamma) = (1-\Gamma)\, 
\xi_{\{0\}}(\rho) + \Gamma\, \xi_{\{1\}}(\rho)$ for linear limb darkening
as a function of the fractional stellar radius $\rho$, where
the two extreme cases of uniform brightness and maximal limb darkening,
corresponding to $\Gamma = 0$ or $\Gamma = 1$, respectively, are shown.}
\label{fig:intprofiles}
\end{figure}

As an example of a parametrized stellar brightness profile, let us
consider the case of linear limb darkening characterized by the
coefficient $0 \leq \Gamma \leq 1$, where
\begin{equation}
\xi(\rho; \Gamma) = (1-\Gamma)\, 
\xi_{\{0\}}(\rho) + \Gamma\, 
\xi_{\{1\}}(\rho)\label{eq:linld}\,,
\end{equation}
with the basis functions
\begin{eqnarray}
\xi_{\{0\}}(\rho)  & = & 1\,,\\ 
\xi_{\{1\}}(\rho) & = & \frac{3}{2}\,\sqrt{1-\rho^2}\,.
\end{eqnarray}
While $\xi_{\{0\}}(\rho)$ corresponds 
to a uniformly bright source, $\xi_{\{1\}}(\rho)$ is linear in
$\cos \vartheta = \sqrt{1-\rho^2}$, where $\vartheta$ denotes the
emergent angle.
These basis brightness profile functions 
 are displayed in
Fig.~\ref{fig:intprofiles}. Any profile $\xi(\rho; \Gamma)$
of the form given by 
Eq.~(\ref{eq:linld}) is a linear superposition of these curves and falls
between them.

\begin{figure}
\includegraphics[width=84mm]{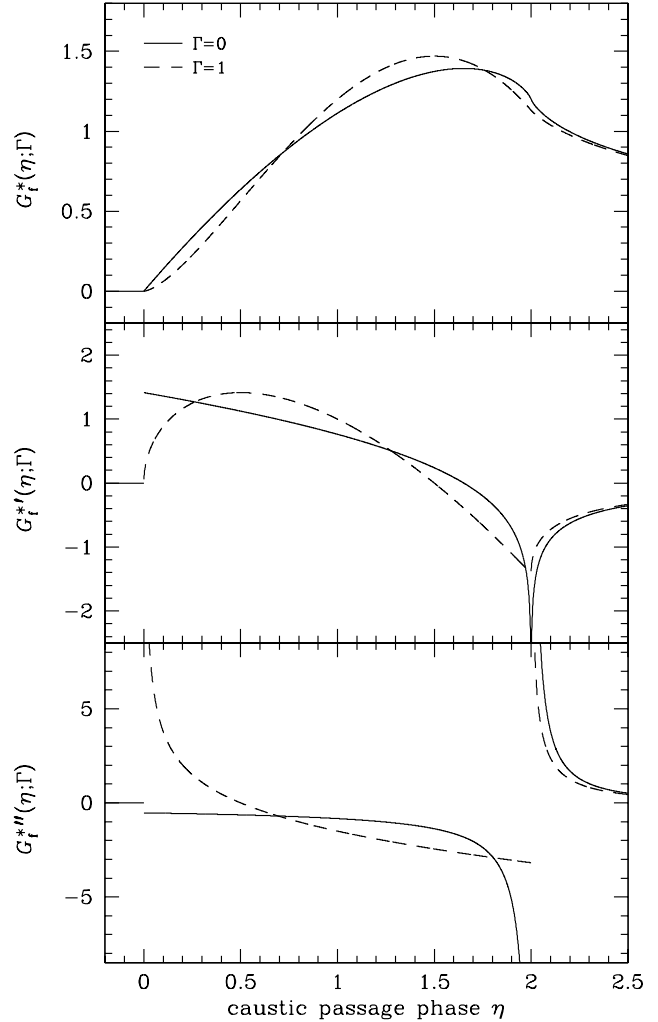}
\caption{The caustic profile function $G_\rmn{f}^\star(\eta; \Gamma) =
(1-\Gamma)\,G_\rmn{f,\{0\}}^\star(\eta) + \Gamma\,
G_\rmn{f,\{1\}}^\star(\eta)$ and its derivatives ${G_\rmn{f}^\star}'(\eta; \Gamma)$ 
and ${G_\rmn{f}^\star}''(\eta; \Gamma)$ as a function of the caustic
passage phase $\eta$. Solid lines correspond to uniform brightness
($\Gamma = 0$) and dashed lines to maximal limb darkening ($\Gamma = 1$), while
general profiles with $0 < \Gamma < 1$ fall between these two extreme case.}
\label{fig:caustprofile}
\end{figure}

\begin{figure}
\includegraphics[width=84mm]{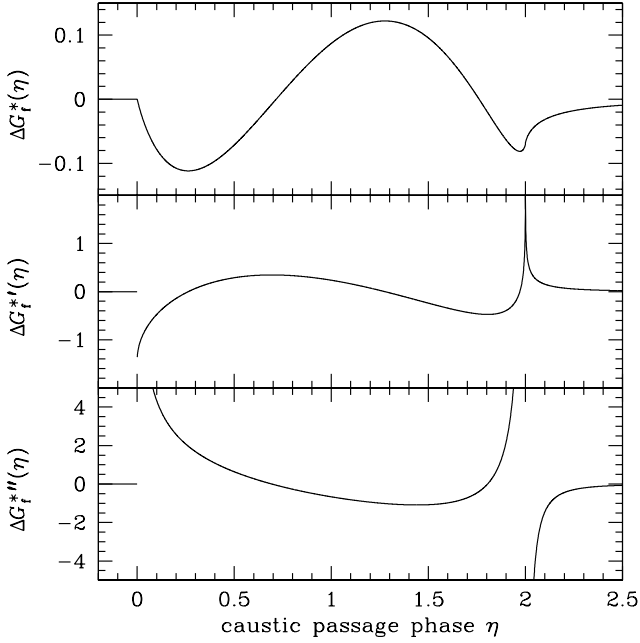}
\caption{Difference between profile functions 
$\Delta G_\rmn{f}^\star(\eta) = G_\rmn{f,\{1\}}^\star(\eta) - G_\rmn{f,\{0\}}^\star(\eta)$
and its derivatives $\Delta {G_\rmn{f}^\star}'(\eta)$ and $\Delta {G_\rmn{f}^\star}''(\eta)$
measuring the sensitivity of the lightcurve to the 
limb-darkening coefficient $\Gamma$
as a function of the
caustic passage phase $\eta$.}
\label{fig:caustgamma}
\end{figure}

For such brightness profiles,
the caustic profile function reads
\begin{equation}
G_\rmn{f}^\star(\eta; \Gamma) = (1-\Gamma)\,G_\rmn{f,\{0\}}^\star(\eta)
+ \Gamma\,G_\rmn{f,\{1\}}^\star(\eta)\,,
\end{equation} 
where $G^{\star}_{\rmn{f},\{0\}}(\eta)$ and $G^{\star}_{\rmn{f},\{1\}}(\eta)$
denote the caustic profile functions 
for the basis stellar brightness profile functions,
which can be
expressed by means of analytic or semi-analytic functions as 
\citep{SW1987, SchneiWag, Do:Fold}
\begin{equation}
G^{\star}_{\rmn{f},\{0\}}(\eta) = \left\{\begin{array}{l}
0 \hfill \mbox{for} \quad \eta \leq 0 \\
\hspace*{-0.1em}
\frac{4\,\sqrt{2}}{3 \upi}\left[(2-\eta)\,
K\left(\sqrt{\eta/2}\,\right)\right.- \\  \quad\left. - 2\,(1-\eta)\,
E\left(\sqrt{\eta/2}\,\right)\right] \\
\qquad \qquad \qquad \qquad  \hfill \mbox{for} \quad
 0 < \eta < 2 \\
\hspace*{-0.1em}
\frac{8}{3 \upi}\sqrt{\eta}\left[(2-\eta)\,
K\left(\sqrt{2/\eta}\,\right)\right.- \\ \quad  \left.  -  (1-\eta)\,
E\left(\sqrt{2/\eta}\,\right)\right]  \; \hfill \mbox{for} \quad
 \eta \geq 2 \end{array}\right.\!\!\!\,,
\end{equation}
where $K$ and $E$ denote the complete elliptical integrals of the
first or second kind,
and
\begin{equation}
G^{\star}_{\rmn{f},\{1\}}(\eta) = \left\{\begin{array}{l}
0 \qquad \qquad \hfill \mbox{for} \quad \eta \leq 0 \\
\frac{2}{5}\,(5-2 \eta)\,\eta^{3/2} \qquad \qquad \hfill \mbox{for} \quad
 0 < \eta \leq 2 \\
\frac{2}{5}\,\left[(5-2 \eta)\,\eta^{3/2}\right.\,+ \\
 \quad +\,\left.(1+2 \eta)\,(\eta-2)^{3/2}\right] \;  \hfill \mbox{for} \quad
 \eta > 2 \\
\end{array}\right.\!\!\!\,,
\end{equation}
respectively, and are shown in Fig.~\ref{fig:caustprofile} together with their
derivatives ${G_\rmn{f}^\star}' = \rmn{d} G_\rmn{f}^\star/\rmn{d} \eta$ and
${G_\rmn{f}^\star}'' = \rmn{d}^2 G_\rmn{f}^\star/\rmn{d} \eta^2$. 
For any choice of $\Gamma$, $G_\rmn{f}^\star(\eta; \Gamma)$ 
vanishes for $\eta \leq 0$, and rises to a peak,
which is shifted towards smaller $\eta$ for stronger limb darkening.
Asymptotically, the decrease after the peak behaves as $G_\rmn{f}^\star(\eta; \Gamma) \simeq
(\eta-1)^{-1/2}$ regardless
of the adopted stellar brightness profile, which corresponds to the behaviour of a 
point-like source.
While the slope of $G^{\star}_{\rmn{f},\{1\}}(\eta)$ vanishes for $\eta = 0$ and remains
finite for $\eta = 2$, it shows a jump discontinuity at $\eta = 0$ and a negative
infinite value at $\eta = 2$ for 
all caustic profiles $G_\rmn{f}^\star(\eta; \Gamma)$ with $\Gamma \neq 1$.
The trend of initially smaller values of the profile function $G_\rmn{f}^\star(\eta; \Gamma)$ and
its slope for stronger limb darkening is reversed on the rise to the peak.
The slope shows strong variations for small $\eta > 0$ and near $\eta = 2$, and little
variation in the remaining regions.
For $\eta = 0$, the change of slope with $\eta$ becomes infinite unless $\Gamma = 0$,
for which a finite negative value is taken. Approaching $\eta = 2$ from larger values, 
the slope of ${G_\rmn{f}^\star}(\eta; \Gamma)$ diverges to positive infinity, whereas it diverges
to negative infinity for the approach from smaller values, except for $\Gamma = 1$, where
a finite negative value is approached.

A change in the linear limb-darkening coefficient $\rmn{d}\Gamma$ causes the
brightness profile $\xi(\rho; \Gamma)$ to vary at each fractional radius $\rho$
by $[\partial \xi(\rho; \Gamma)/\partial \Gamma]\,\rmn{d}\Gamma$. Due to the
linearity of $\xi(\eta; \Gamma)$ in $\Gamma$, 
$\partial \xi(\rho; \Gamma)/\partial \Gamma$ equals
the difference between the basis profiles, i.e.\
\begin{equation}
\frac{\partial \xi(\rho; \Gamma)}{\partial \Gamma} = \xi_{\{1\}}(\rho)-
\xi_{\{0\}}(\rho) \equiv \Delta \xi(\rho)\,,
\end{equation}
while all higher derivatives of $\xi(\rho; \Gamma)$ vanish identically.\footnote{
Such a property does not hold for the popular parametrization of linear limb darkening
where $\xi(\rho; c) = (1-c/3)^{-1}\,\left(1-c + c\,\sqrt{1-\rho^2}\right)$.} 
As Fig.~\ref{fig:intprofiles} shows, $\Delta \xi(\rho)$ decreases monotonically from
$\Delta \xi(0) = 1/2$ to $\Delta \xi(1) = -1$. With $\rmn{d}\Delta \xi(\rho)/
\rmn{d}\rho = -(3/2)\,\rho\,(1-\rho^2)^{-3/2}$, its slope becomes flat at the 
source center and decreases monotonically towards the limb where it becomes infinite.
The normalization forces $\Delta \xi(\rho)$ to change it sign, which occurs
for $\rho_0 = \sqrt{5}/3 \approx 0.745$.

Changes in $\xi(\rho; \Gamma)$ caused by variations of $\Gamma$ alter the
caustic profile function $G_\rmn{f}^\star(\eta; \Gamma)$ according to 
Eq.~(\ref{eq:response}) 
with the sensitivity of the caustic profile at a given passage phase $\eta$
to these variations characterized
by ${\bmath{\mathcal T}} (\eta,\rho)$ as discussed in the previous section.
$G_\rmn{f}^\star(\eta; \Gamma)$ inherits the linearity in $\Gamma$ from the brightness profile
$\xi(\rho; \Gamma)$, so that
\begin{equation}
\frac{\partial G_\rmn{f}^\star(\eta; \Gamma)}{\partial \Gamma} =
  G_\rmn{f,\{1\}}^\star(\eta) - G_\rmn{f,\{0\}}^\star(\eta) \equiv 
\Delta G_\rmn{f}^\star(\eta)\,,
\end{equation}
measures the sensitivity of the lightcurve to variations of the limb-darkening coefficient
$\Gamma$ and for its derivatives, one finds in analogy 
$\partial {G_\rmn{f}^\star}'(\eta; \Gamma)/\partial \Gamma = {\Delta G_\rmn{f}^\star}'(\eta)$ and
$\partial {G_\rmn{f}^\star}''(\eta; \Gamma)/\partial \Gamma = {\Delta G_\rmn{f}^\star}''(\eta)$.
These functions are displayed in Fig.~\ref{fig:caustgamma}.

Since the stellar magnitude $m_\rmn{fold}(t)$ is 
a function of elapsed time $t$ rather than of
caustic passage phase $\eta$,
a reliable measurement of the limb-darkening coefficient $\Gamma$ requires not
only a sufficient
sensitivity of the caustic profile 
function $G_\rmn{f}^\star(\eta; \Gamma)$ to the variation of $\Gamma$, but
in addition also the determination of the parameters $t_\rmn{f}^\star$,
$t_\star^\perp$, $m_\rmn{f}^\star$, $g_\rmn{f}^\star$, and $\hat \omega_\rmn{f}^\star$,
which establish the
relation between $G_\rmn{f}^\star(\eta: \Gamma)$ and $m_\rmn{fold}(t)$, as pointed out in the
previous section. 
The determination of these parameters
from the photometric data over a given region of the lightcurve can be viewed
in relation to the local assessment of the caustic profile function
$G_\rmn{f}^\star(\eta; \Gamma)$ and its derivatives ${G_\rmn{f}^\star}'(\eta; \Gamma)$ and
${G_\rmn{f}^\star}''(\eta; \Gamma)$ with respect to the caustic passage phase $\eta$
as well as their derivatives with respect to $\Gamma$, namely
$\Delta G_\rmn{f}^\star(\eta)$, ${\Delta G_\rmn{f}^\star}'(\eta)$, and 
${\Delta G_\rmn{f}^\star}''(\eta)$, where the amount of information provided increases with
the absolute value of each of these quantities.

With ${\bmath{\mathcal T}} (\eta,\rho) \geq 0$, the sign change in $\Delta \xi$ at
$\rho_0 = \sqrt{5}/3$ causes $\Delta G_\rmn{f}^\star(\eta)$ to change its sign at
$\eta_0^\rmn{out} = 0.716$ and
$\eta_0^\rmn{in} = 1.762$, where $\Delta G_\rmn{f}^\star(\eta) < 0$ for
$0 < \eta < \eta_0^\rmn{out}$ or $\eta > \eta_0^\rmn{in}$, whereas 
$\Delta G_\rmn{f}^\star(\eta) > 0$ for $\eta_0^\rmn{out} < \eta < \eta_0^\rmn{in}$.
A change in $\Gamma$ causes the largest relative changes
in the observed flux at the extrema of $\Delta G_\rmn{f}^\star$, which occur at
$\eta_\rmn{min}^\rmn{out} = 0.263$, $\eta_\rmn{max} = 1.276$, and $\eta_\rmn{min}^\rmn{in} = 1.971$,
i.e.\ close to the point at which the caustic is touched by the limb of the source
located inside.

By means of the sensitivity function ${\bmath{\mathcal T}} (\eta,\rho)$, 
variations of $\eta$ during the caustic passage alter the weight given to the
different fractional
radii $\rho$ and $|\Delta {G_\rmn{f}^\star}'(\eta)|$ 
increases with the absolute variation in $\Delta \xi(\rho)$ for the specific radii that
dominate ${\bmath{\mathcal T}} (\eta,\rho)$ for a given caustic passage phase $\eta$.
With $\rmn{d}\Delta \xi(\rho)/
\rmn{d}\rho|\Delta \xi(\rho) < 0$ decreasing monotonically
from the souce center to the limb, and
moreover, the caustic producing a weaker response characterized by
${\bmath{\mathcal T}} (\eta,\rho)$ to variations of the brightness profile near the source center,
the variations in $\Delta G_\rmn{f}^\star(\eta)$ are small for $\eta \sim 1$ and large
as $\eta \to 0$ or $\eta \to 2$. In fact, Fig.~\ref{fig:caustgamma} shows that 
$|\Delta {G_\rmn{f}^\star}'(\eta)|$ assumes large values for small positive $\eta$ and in
the close vicinity of $\eta \sim 2$, and becomes infinite at $\eta = 0$ and $\eta = 2$,
whereas much smaller values are assumed for the remaining part of the caustic passage, and
the variation of $\Delta {G_\rmn{f}^\star}'(\eta)$, given by $\Delta {G_\rmn{f}^\star}''(\eta)$,
behaves in a similar way.
Roughly speaking, when comparing different phases of the caustic passage, 
the largest response to a variation of the limb-darkening coefficient
is caused around the 
time when the stellar limb touches the caustic with the source center being inside 
($\eta \sim 2$) 
with all $\Delta {G_\rmn{f}^\star}(\eta)$, $\Delta {G_\rmn{f}^\star}'(\eta)$,
and $\Delta {G_\rmn{f}^\star}''(\eta)$ assuming reasonably large absolute values or even tending to
infinity as $\eta \to 2$. Next comes the region over which
the stellar limb touches the caustic while the
source center is outside, where in contrast $|\Delta {G_\rmn{f}^\star}'(\eta)|$ decreases to zero 
as positive $\eta \to 0$ and vanishes identically for $\eta \leq 0$. 
In contrast, much smaller effects result
while the inner parts of the source pass the caustic, for which 
both $\Delta {G_\rmn{f}^\star}'(\eta)$ and $\Delta {G_\rmn{f}^\star}''(\eta)$ are
close to zero or $|\Delta {G_\rmn{f}^\star}(\eta)|$ is small
if one of the former quantities
tends to assume a reasonable non-zero value.

In order to assess the potential of different regions of an observed lightcurve
for the measurement of the (linear) limb-darkening coefficient $\Gamma$, one
needs to consider not only the flux variation with $\Gamma$ at the corresponding
caustic phase, but also the observed flux and its
uncertainty. Let $F_\rmn{f}^\star$ denote the flux at time
$t_\rmn{f}^\star$, so that
$m_\rmn{fold}(t) = m_\rmn{f}^\star = -2.5 \lg [F_\rmn{fold}(t)/
F_\rmn{f}^\star]$. 
The ratio $Z(\eta)$ between signal-to-noise for flux deviations 
caused by a variation of the limb-darkening coefficient $\Gamma$ 
and signal-to-noise for the flux $F_\rmn{f}^\star$ is given by
\begin{equation}
Z(\eta) = \frac{\partial F_\rmn{fold}\left(t_\rmn{f}^\star \pm \eta\,
t_\star^\perp; \Gamma\right)}
{\partial \Gamma}\,\frac{\sigma_\rmn{f}^\star}{F_\rmn{f}^\star\,\sigma}\,.
\end{equation}
With the assumption that the flux uncertainties $\sigma$ follow
Poisson statistics, so that $\sigma = \sigma_\rmn{f}^\star\,
\left(F/F_\rmn{f}^\star\right)^{1/2}$, where $\sigma_\rmn{f}^\star$ is
the uncertainty of $F_\rmn{f}^\star$, and neglecting the temporal flux variation
in non-critical images (i.e.\ adopting $\hat \omega_\rmn{f}^\star = 0$),
one obtains 
\begin{equation}
Z(\eta) =
\alpha_\Gamma
\frac{\Delta G^{\star}_{\rmn{f}}(\eta)}
{\sqrt{\alpha_\Gamma\,G^{\star}_{\rmn{f}
}(\eta; \Gamma) +
1}}\,,
\end{equation}
where
\begin{equation}
\alpha_\Gamma = \frac{(F_{\rmn{peak}}/F_\rmn{f}^\star) -1}
{G_{\rmn{peak},\Gamma}}
\end{equation}
and $G_{\rmn{peak},\Gamma}$ is the maximum of
$G^{\star}_{\rmn{f}}(\eta; \Gamma)$ which corresponds to the
observed flux $F_{\rmn{peak}}$.

\begin{figure}
\includegraphics[width=84mm]{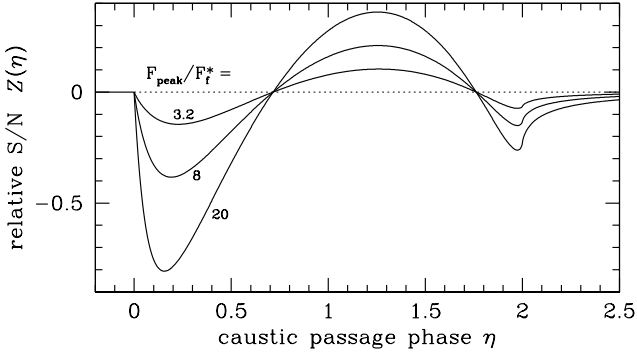}
\caption{Ratio $Z(\eta)$ between signal-to-noise for the
flux variation $[\partial F(t_\rmn{f}^\star+\eta\,t_\star^\perp; \Gamma)/
\partial \Gamma]/\sigma$ caused by altering the
limb-darkening coefficient $\Gamma$ and
signal-to-noise $F_\rmn{f}^\star/\sigma_\rmn{f}$ for the
flux measurement outside the caustic for selected caustic-peak-to-outside
magnification ratios
$F_\rmn{peak}/F_\rmn{f}^\star$.}
\label{fig:zofeta}
\end{figure}

Using $\Gamma = 1$ as reference, $Z(\eta)$ is shown in Fig.~\ref{fig:zofeta}
for selected values of the peak-to-outside flux ratio 
$F_\rmn{peak}/F_\rmn{f}^\star$.
Compared to the effects on the flux difference 
as displayed in Fig.~\ref{fig:caustgamma},
caustic passage phases near the
beginning of an entry or the end of an exit are favoured, yielding a
maximal signal for phases close to -- but not at -- this point. However,
$|{\Delta G_\rmn{f}^\star}'(\eta)|$ is small in this 
region, whereas
it peaks for $\eta = 2$, i.e.\ when the caustic entry is completed or the
caustic exit starts. 

For $Z(\eta) = 20\,(\sigma_\rmn{f}^\star/F_\rmn{f}^\star)$,
a single data point implies an absolute uncertainty of $0.05$ for the
limb-darkening coefficient $\Gamma$. If one considers $F_\rmn{peak}/F_\rmn{f}^\star \sim 8$ as for recently observed events, Fig.~\ref{fig:zofeta} yields
$|Z(\eta)| \sim 0.25$, so that for a photometric precision of a few percent
at the caustic outside, tens of data points yield a reliable measurement
of $\Gamma$.

The identification and assessment of some
distinctive characteristic local features in the light curve that
do not depend on the stellar brightness profile allow 
the precise determination of individual lightcurve parameters, leading
to a reduction in parameter degeneracies and uncertainties. 
For any stellar brightness profile that does not vanish at the stellar limb,
the slope of the lightcurve exhibits a jump discontinuity at $\eta = 0$ and
its identification in the lightcurve directly yields
$m_\rmn{f}^\star$ and $t_\rmn{f}^\star$. 
Another distinctive feature for such brightness profiles
is the infinite slope and sign change of
the curvature  
at $\eta =2$, where temporal identification of both $\eta = 0$
and $\eta = 2$ yield $t_\star^\perp$.
With $G_\rmn{f}^\star(\eta; \Gamma) \equiv 0$ for $\eta < 0$, data taken
while the source is completely outside the caustic provide $\hat \omega_\rmn{f}$, and
with $m_\rmn{f}^\star$ and 
$g_\rmn{f}^\star$ can then be measured from
the asymptotic form $G_\rmn{f}^\star(\eta; \Gamma) \simeq (\eta-1)^{-1/2}$ for 
sources inside the caustic.
The possibility of measuring each of the parameters 
$t_\rmn{f}^\star$, $m_\rmn{f}^\star$, and
$\hat \omega_\rmn{f}$ directly from data obtained around the passage
of the stellar limb over the caustic while the source is 
outside ($\eta \sim 0$) makes this phase of the caustic passage
more valuable than that of the limb passage while the source is inside
($\eta \sim 2$), which alone does not provide 
any direct measurements of individual light curve parameters. Only the combination
of both of these phases yields direct measurements of $t_\star^\perp$ and $g_\rmn{f}^\star$.
In contrast, data taken while the inner parts of the source pass the caustic ($\eta \sim 1$)
do not provide any direct measurements of lightcurve parameters neither alone nor
in combination with data in other regions.

\section{Simulated photometric data}
\label{sec:sampled}
\subsection{Creation of dataset}
In order to study the uncertainties of the measurement of the linear limb-darkening coefficient
in dependence of the sampling of the lightcurve for different regions, a simulated dataset
for a typical caustic exit 
in accordance with the capabilities of the PLANET campaign \citep{PLANET:EGS} has been
created. Contrary to caustic entries, which have to be caught by chance, 
caustic exits can be predicted and therefore provide regular opportunities
to obtain high-quality data during caustic passages. 
For a G/K subgiant in the Galactic bulge at a distance $D_\rmn{S} \sim
8.5~\mbox{kpc}$  with a
brightness $I \sim 17$ and a radius $R \sim 3~R_\star$,
the angular radius becomes $\theta_\star \sim 2~\mbox{$\mu$as}$, so 
that the choice of $t_\rmn{f}^\star = 6~\mbox{h}$ 
corresponds to a proper motion perpendicular to the
caustic of $\mu^\perp \sim 7~\mbox{$\mu$as}\,\mbox{d}^{-1}$ and to a velocity 
$v^\perp \sim 70~\mbox{km}\,\mbox{s}^{-1}$ at the lens distance
$D_\rmn{L} \sim 6~\mbox{kpc}$.
For the unit time ${\hat t} = 1~\mbox{h}$ and the adopted
limb-darkening coefficient $\Gamma = 0.5$,
the choice $g_\rmn{f}^\star = 0.05$ yields
a relative magnification $A_\rmn{peak}/A_\rmn{f}^\star \sim 12.5$ of the caustic
peak relative to the caustic exit (at time $t_\rmn{f}^\star$), for which we assume 
$m_\rmn{f}^\star =16.3$.
Finally, ${\hat \omega}_\rmn{f}^\star = -0.001~\mbox{h}^{-1}$ and 
$t_\rmn{f}^\star = 0$ are adopted.

\begin{figure}
\includegraphics[width=84mm]{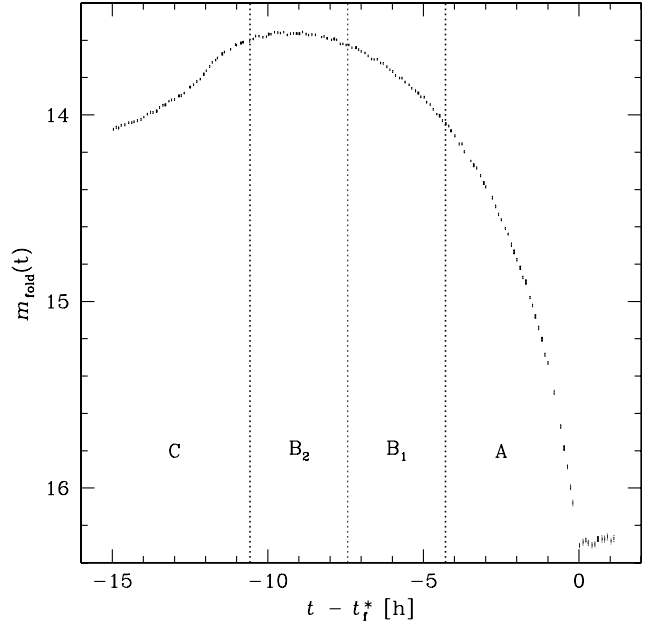}
\caption{Simulated dataset for the lightcurve parameters
$t_\rmn{f}^\star = 0$,
$t_\star^\perp = 6~\mbox{h}$, $m_\rmn{f}^\star = 16.3$,
$g_\rmn{f}^\star = 0.05$ (${\hat t} = 1~\mbox{h}$), 
${\hat \omega}_\rmn{f}^\star = -0.001~\mbox{h}^{-1}$, and $\Gamma = 0.5$.
The standard sampling interval is $\Delta t = 6~\mbox{min}$, while data corresponding 
to a sparser sampling with $\Delta t = 30~\mbox{min}$ only are displayed
in bold. The measurement uncertainty has been assumed to be 0.015~mag
for $m_\rmn{f}^\star$ and scaled to other magnitudes assuming the flux to follow Poisson
statistics. More details can be found in the text.
The division of the dataset into an outside region A, a central region B,
with subregions B$_1$ and B$_2$, and an inside region C, is indicated
by vertical dashed lines.}
\label{fig:OB66005}
\end{figure}

The sampling is characterized by the sampling interval
$\Delta t = 6~\mbox{min}$, an uncertainty $\sigma_{\Delta t} = 45~s$ 
of the point of time at which 
the measurement is taken, 
and the sampling phase shift $t_\rmn{phase} = 1.2~\mbox{min}$.
The uncertainties in the observed magnitudes have been assumed to be $\sigma_0 = 0.015$
at $m_\rmn{f}^\star$ (i.e.\ the relative uncertainty of the measured flux
is $\sim\,1.5\,$\%) and scaled to other magnitudes, so that the uncertainty for the
corresponding flux follows Poisson statistics in being
proportional to the square-root of its value. Moreover, the
uncertainty of the magnitude at given time has been smeared with a relative 
standard deviation of $f_\sigma = 0.125$.

For $t_\rmn{min}Ê= -15~\mbox{h}$ and $t_\rmn{max} = t_\rmn{min}+n\,\Delta t = 1.2~\mbox{h}$,
a synthetic dataset ($t^{(i)}$, $m^{(i)}$, $\sigma^{(i)}$) containing $n = 162$ simulated
measurements of the
time of observation, the magnitude, and its uncertainty 
has thus been created using
\begin{eqnarray} 
t^{(i)} & = & \mathcal{N}\left(t_\rmn{min}+i\,\Delta t +t_\rmn{phase},\sigma_{\Delta t}\right)\,,\\
\sigma^{(i)} & = & \mathcal{N}\left( 
\sigma_0\,10^{\left\{0.2\,\left[m_\mathrm{f}^\star-m_\rmn{fold}(t^{(i)})\right]\right\}},\right. \nonumber \\
& & \quad \left.f_{\sigma}\,\sigma_0\,10^{\left\{0.2\,\left[m_\mathrm{f}^\star-m_\rmn{fold}(t^{(i)})\right]\right\}}
\right)\,,\label{eq:scalesigma}
\\
m^{(i)} & = & \mathcal{N}\left(m_\rmn{fold}(t^{(i)}),\sigma^{(i)}\right)\,,
\end{eqnarray}
where $\mathcal{N}(\mu,\sigma)$ denotes a value drawn randomly from a normal distribution with mean
$\mu$ and standard deviation $\sigma$.

In order to simulate possible losses in the data acquisition 
due to observing conditions
or technical problems, data points have been removed from the
dataset at random with a probability of $q_\rmn{l} = 5\,$\%.
By random selection of data points with a probability of
$q_\rmn{s} = 20~\,$\%, a restricted dataset has been obtained,
which simulates a sparser sampling
with an expected average sampling interval of 30~min.
The data used for the subsequent analysis are shown in
Fig.~\ref{fig:OB66005}, with 27 points, which 
belong to the sparse sampling only
shown in bold,
while the dense sampling includes
all 153 points shown in bold and in light.

Missing or sparse coverage over part of a caustic passage are likely to occur if
its duration exceeds the time of coverage possible with a single telescope.
Although the coverage by more than one telescope would require different parameters
$m_\rmn{f}^{\star\,(s)}$ and $g_\rmn{f}^{\star\,(s)}$ \citep[c.f.][]{Do:Fold},
a careful assessment of parallax between the sites \citep{PLANET:EB5mass}, and
different achievable photometric precisions, this will
be neglected in the following 
for reasons of simplicity, where site-independent fit parameters $m_\rmn{f}^\star$
and $g_\rmn{f}^\star$ will be used, which is equivalent to treating the data as resulting
from observations with a single telescope.

\subsection{Tests on statistical properties}
\label{sec:stattests}

In order to provide a reference and to check the created data set, 
let us have a look some of its statistics 
with respect to the underlying set of true parameters $\vec p_\rmn{true}$.
 
Let us follow \citet{Rhie:LD} in using the caustic passage phases at which 
$\Delta G_\rmn{f}^\star(\eta)$ vanishes, namely $\eta_0^\rmn{out} = 0.716$ and
$\eta_0^\rmn{in} = 1.762$ as boundaries to divide the data into an
outside region A ($\eta < \eta_0^\rmn{out}$), a central region B
($\eta_0^\rmn{out} \leq \eta < \eta_0^\rmn{in}$), and an inside region
C ($\eta \geq \eta_0^\rmn{in}$). Furthermore, let us subdivide the central
region B at $\eta_0^\rmn{c} = (\eta_0^\rmn{out} + \eta_0^\rmn{in})/2$
into two regions B$_1$ ($\eta_0^\rmn{out} \leq  \eta < \eta_0^\rmn{c}$)
and B$_2$ ($\eta_0^\rmn{c} \leq \eta < \eta_0^\rmn{in}$).
These regions are indicated in Fig.~\ref{fig:OB66005}.  

\begin{table}
\caption{Number of data points $N$ and scatter of residuals for different
regions of the simulated dataset.}
\label{tab:datasetprop}
\begin{tabular}{@{}ccccc}
\hline
Region & $\chi^2_\rmn{true}$ & $N$ & $\chi^2_\rmn{true}/N$ &
$P_{\chi^2_\rmn{true}}$   
 \\
\hline
A & 51.5 & 50 & 1.02 & 0.42 \\
B$_1$ & 34.5 & 31 & 1.11 & 0.30 \\
B$_2$ & 42.8 & 30 & 1.43 & 0.06 \\
C & 36.7 & 42 & 0.87 & 0.67 \\
ABC & 165.5 & 153 & 1.08 & 0.23 \\
a & 5.69 & 11 & 0.52 & 0.89 \\
b$_1$ & 3.40 & 3 & 1.13 & 0.33 \\
b$_2$ & 4.38 & 5 & 0.88 & 0.50 \\
c & 2.70 & 8 & 0.34 & 0.95 \\
abc & 16.17 & 27 & 0.60 & 0.95 \\
\hline
\end{tabular}

\medskip
$\chi^2_\rmn{true}
= \chi^2(\vec p_\rmn{true})$ is the value of $\chi^2$ for the
true parameters $\vec p_\rmn{true}$ and 
$P_{\chi^2_\rmn{true}}
= P(\chi^2 \geq \chi^2_\rmn{true})$.
Uppper-case letters refer to the dense and lower-case letters to
the sparse sampling. 
\end{table}

\begin{table*}
\begin{minipage}{176mm}
\caption{Run tests on the lightcurve residuals for the true
model parameters.}
\label{tab:runtesttrue}
\begin{tabular}{@{}ccccccccc}
\hline
 Region & 
$N$ & $N_{+}$ & $N_{-}$ & $\mathcal{E}(n_\rmn{r})$ & 
$\sigma(n_\rmn{r})$ & $n_\rmn{r}^\rmn{obs}$ 
& $\delta$ & $P_\rmn{r}$ \\ 
\hline
ABC &
153 & 68 & 85 & 76.56 & 6.09 & 75 & 0.26 & 0.40\\ 
abc &
27 & 14 & 13 & 14.48 & 2.54 & 12 & 0.98 & 0.16 \\ 
\hline
\end{tabular}

\medskip
From the total number of $N$ data points,
$N_{+}$ show positive and $N_{-}$ show negative residuals, 
implying $\mathcal{E}(n_\rmn{r})$ runs
with a standard deviation $\sigma(n_\rmn{r})$ to be expected. With 
$n_\rmn{r}^{\rmn{obs}}$ observed runs, the deviation
$\delta = [\mathcal{E}(n_\rmn{r})-n_\rmn{r}^{\rmn{obs}}]/
\sigma(n_\rmn{r})$ is related to a probability $P_\rmn{r} = P(n_\rmn{r} \leq 
n_\rmn{r}^\rmn{obs})$. 
\end{minipage}
\end{table*}

Table~\ref{tab:datasetprop} gives the number of data points $N$ for the
full dataset and each of the regions A, B$_1$, B$_2$, and C for
both samplings as well as $\chi^2_\rmn{true} = 
\chi^2(\vec p_\rmn{true})$, 
the ratio $\chi^2_\rmn{true}/N$, 
which is expected to be unity,
and the probability $P(\chi^2 \geq \chi^2_\rmn{true})$.
While the standard upper-case letters refer to the simulated data 
for the dense sampling with $\Delta t = 6~\mbox{min}$,
the sparse sampling for a certain region is indicated by the 
corresponding lower-case letter.
One sees that the average absolute size of the residuals differs among the
different regions, in particular region B$_2$ shows rather large
 residuals while the adjoining region C partly
 compensates this with smaller than
expected residuals. At first sight, it might be surprising to find
a probability of only about 6\,\% for region B$_1$. However, this means
that the probability that at least one of four regions shows a probability
of that order or less is $\sim\,25\%$, so that about a quarter of all
simulated datasets should share this feature, which is reflected in the
probability for the full dataset ABC.
The fraction of points selected for the sparse sampling is
$\sim\,18\,$\%, somewhat smaller than the expected 20\,\%. In particular,
only $\sim\,10\,$\% of the points in region B$_2$ made it to b$_2$.
Regions $a$ and $c$ show residuals whose absolute values 
are on average much smaller 
than expected, with a probability of being smaller than this between
5 and 10\,\%. Two regions with probabilities of about 10~\% make the 
sparse sample less likely than the dense one with 
$P(\chi^2 \leq \chi^2_\rmn{true}) \sim  5\,$\%.

Systematic deviations in the data
can be quantitatively assessed by means of a run test, which
is based on the statistics of the signs of the residuals 
irrespective of their absolute values, whereas a $\chi^2$ test is based
on the absolute values while being blind to the signs. 
Let a 'run' being defined as the longest contiguous sequence of
residuals with the same sign, and let $N$ denote the total number of 
data points, $N_{+}$ the number of points with positive residuals, and
$N_{-}$ the number of points with negative residuals.
For $N > 10$, the distribution of the number of runs $n_\rmn{r}$ can 
be fairly approximated by a normal distribution with the expectation value
\begin{equation}
\mathcal{E}(n_\rmn{r})  =  1+\frac{2 N_{+} N_{-}}{N}
\end{equation}
and the standard deviation
\begin{equation}
\sigma(n_\rmn{r})  =  \sqrt{2 N_{+} N_{-} 
\frac{2 N_{+} N_{-} - N}{N^2 (N-1)}}\,.
\end{equation}
Results of run tests on the residuals of data in all regions for both adopted samplings with respect
to the true lightcurve
are listed in Table~\ref{tab:runtesttrue}, which do not contradict the assumption
that the scatter in the residuals
is random.

Without knowledge of the true model parameters, the assessment of the
goodness-of-fit by means of $\chi^2$ is hampered by the fact that
the reported error bars frequently do not 
properly reflect
the true error bars of the measurement 
\citep[e.g.][]{OGLEerr,PLANET:first} 
forcing
the introduction of modelled scaling factors and/or systematic
errors \citep[e.g][]{EROSerrors,Do:Thesis,Tsapras}, so that the hypothesis of
underestimated error bars rather stands against rejecting the model on a basis
of a larger than expected $\chi^2_\rmn{min}$. This is not a problem for the
assessment of the run test, which does not depend on the size of the error
bars.

\section{Measurement of parameters}
\label{sec:reconstruction}

\begin{table*}
\begin{minipage}{176mm}
\caption{Best-fit parameters and their uncertainties for the simulated data
in selected regions of the lightcurve.}
\label{tab:simu}
\begin{tabular}{@{}ccccccc}
\hline
Region &  $t_\rmn{f}^\star$ [h]& 
$t_\star^\perp$ [h]& $m_\rmn{f}^\star$ &
$g_\rmn{f}^\star$ & $-{\hat \omega}_\rmn{f}^\star$ [h$^{-1}$] & $\Gamma$ \\
\hline
ABC & 
$-0.0001_{-0.0088}^{+0.0091}$ &
$6.0006_{-0.0095}^{+0.0097}$ &
$16.2967_{-0.0040}^{+0.0040}$ &
$0.05021_{-0.00025}^{+0.00025}$ &
$0.00099_{-0.00011}^{+0.00011}$ &
$0.512_{-0.012}^{+0.012}$ \\[0.8ex]
aBC &
$-0.007_{-0.016}^{+0.017}$ &
$5.997_{-0.016}^{+0.017}$ &
$16.287_{-0.011}^{+0.011}$ &
$0.05072_{-0.00060}^{+0.00060}$ &
$0.00099_{-0.00011}^{+0.00012}$ &
$0.512_{-0.020}^{+0.021}$ \\[0.8ex]
Abc &  
$-0.007_{-0.010}^{+0.010}$ &
$6.010_{-0.013}^{+0.013}$ &
$16.2998_{-0.0051}^{+0.0052}$ &
$0.04975_{-0.00048}^{+0.00048}$ &
$0.00128_{-0.00032}^{+0.00032}$ &
$0.522_{-0.016}^{+0.016}$ \\[0.8ex]
aBc &  
$0.00_{-0.02}^{+0.020}$ &
$6.007_{-0.024}^{+0.023}$ &
$16.290_{-0.013}^{+0.013}$ &
$0.05038_{-0.00082}^{+0.00082}$ &
$0.00120_{-0.00037}^{+0.00041}$ &
$0.516_{-0.024}^{+0.024}$ \\[0.8ex]
abC &  
$-0.004_{-0.019}^{+0.020}$ &
$5.987_{-0.017}^{+0.019}$ &
$16.288_{-0.011}^{+0.011}$ &
$0.05060_{-0.00060}^{+0.00061}$ &
$0.00103_{-0.00012}^{+0.00012}$ &
$0.518_{-0.025}^{+0.025}$ \\[0.8ex]
abc & 
$0.000_{-0.023}^{+0.024}$ &
$6.007_{-0.025}^{+0.025}$ &
$16.291_{-0.013}^{+0.013}$ &
$0.05020_{-0.00085}^{+0.00085}$ &
$0.00129_{-0.00040}^{+0.00045}$ &
$0.520_{-0.030}^{+0.031}$ \\[0.8ex]
AB$_1$ & 
$-0.003_{-0.011}^{+0.012}$ &
$6.036_{-0.042}^{+0.044}$ &
$16.3002_{-0.0068}^{+0.0067}$ &
$0.04941_{-0.00082}^{+0.00086}$ &
$0.00130_{-0.00052}^{+0.00051}$ &
$0.499_{-0.019}^{+0.020}$ \\[0.8ex]
AB & 
$0.003_{-0.010}^{+0.010}$ &
$6.004_{-0.016}^{+0.015}$ &
$16.3002_{-0.0068}^{+0.0067}$ &
$0.04981_{-0.00066}^{+0.00072}$ &
$0.00133_{-0.00053}^{+0.00052}$ &
$0.512_{-0.013}^{+0.013}$ \\[0.8ex]

Ab &
$0.009_{-0.012}^{+0.010}$ &
$5.998_{-0.021}^{+0.021}$ &
$16.3001_{-0.0067}^{+0.0066}$ &
$0.04984_{-0.00067}^{+0.00073}$ &
$0.00133_{-0.00053}^{+0.00052}$ &
$0.529_{-0.018}^{+0.018}$ \\[0.8ex]
aB &
$0.062_{-0.060}^{+0.099}$ &
$5.989_{-0.049}^{+0.054}$ &
$16.41_{-0.11}^{+0.13}$ &
$0.0422_{-0.0052}^{+0.0066}$ &
$0.001_{-0.008}^{+0.010}$ &
$0.430_{-0.092}^{+0.080}$ \\[0.8ex]
ab &
$0.07_{-0.17}^{+0.09}$ &
$5.975_{-0.046}^{+0.067}$ &
$16.43_{-0.23}^{+0.12}$ &
$0.0418_{-0.005}^{+0.028}$ &
$0.011_{-0.018}^{+0.010}$ &
$0.44_{-0.09}^{+0.10}$ \\[0.8ex]
AC & 
$0.002_{-0.011}^{+0.011}$ &
$6.001_{-0.010}^{+0.011}$ &
$16.2970_{-0.0040}^{+0.0040}$ &
$0.05013_{-0.00027}^{+0.00027}$ &
$0.00102_{-0.00012}^{+0.00013}$ &
$0.517_{-0.020}^{+0.020}$ \\[0.8ex]
Ac &
$0.006_{-0.012}^{+0.012}$ &
$6.011_{-0.014}^{+0.013}$ &
$16.3000_{-0.0051}^{+0.0052}$ &
$0.04970_{-0.00049}^{+0.00049}$ &
$0.00132_{-0.00033}^{+0.00034}$ &
$0.521_{-0.020}^{+0.021}$ \\[0.8ex]
aC &
$-0.023_{-0.021}^{+0.024}$ &
$5.984_{-0.017}^{+0.020}$ &
$16.287_{-0.012}^{+0.012}$ &
$0.05082_{-0.00063}^{+0.00064}$ &
$0.00095_{-0.00013}^{+0.00014}$ &
$0.482_{-0.036}^{+0.039}$ \\[0.8ex]
ac &
$-0.014_{-0.026}^{+0.028}$ &
$5.998_{-0.026}^{+0.026}$ &
$16.291_{-0.013}^{+0.013}$ &
$0.05027_{-0.00089}^{+0.00088}$ &
$0.00128_{-0.00041}^{+0.00047}$ &
$0.493_{-0.042}^{+0.044}$ \\[0.8ex]
B$_2$C & 
$0.19_{-0.61}^{+0.34}$ &
$6.10_{-0.31}^{+0.17}$ &
$16.7_{-1.2}^{+2.7}$ &
$0.034_{-0.032}^{+0.082}$ &
$0.00032_{-0.0008}^{+0.0045}$ &
$0.494_{-0.048}^{+0.054}$ \\[0.8ex]
BC & 
$0.09_{-0.46}^{+0.31}$ &
$6.05_{-0.23}^{+0.16}$ &
$16.6_{-0.9}^{+2.3}$ &
$0.039_{-0.035}^{+0.060}$ &
$0.0004_{-0.0009}^{+0.0032}$ &
$0.523_{-0.028}^{+0.032}$ \\[0.8ex]
Bc &
$-0.31_{-0.54}^{+0.70}$ &
$5.86_{-0.27}^{+0.35}$ &
$15.7_{-0.6}^{+2.1}$ &
$0.095_{-0.085}^{+0.081}$ &
$0.0038_{-0.0042}^{+0.0052}$ &
$0.562_{-0.049}^{+0.052}$ \\[0.8ex]
bC &
$0.25_{-0.47}^{+0.22}$ &
$6.13_{-0.24}^{+0.12}$ &
$17.1_{-1.3}^{+2.1}$ &
$0.024_{-0.022}^{+0.059}$ &
$-0.0001_{-0.0004}^{+0.0029}$ &
$0.521_{-0.036}^{+0.038}$ \\[0.8ex]
bc &
$0.1_{-1.1}^{+0.3}$ &
$6.08_{-0.53}^{+0.20}$ &
$16.5_{-1.5}^{+2.3}$ &
$0.04_{-0.04}^{+0.15}$ &
$0.0008_{-0.0013}^{+0.0095}$ &
$0.545_{-0.052}^{+0.063}$ \\[0.8ex]
A & 
$-0.010_{-0.015}^{+0.016}$ &
$6.66_{-0.49}^{+0.67}$ &
$16.3004_{-0.0069}^{+0.0068}$ &
$0.0425_{-0.0058}^{+0.0053}$ &
$0.00112_{-0.00046}^{+0.00048}$ &
$0.466_{-0.043}^{+0.044}$ \\[0.8ex]
B & 
$-0.3_{-1.1}^{+1.2}$ &
$6.0_{-1.5}^{+0.8}$ &
$15.3_{-0.5}^{+2.0}$ &
$0.15_{-0.14}^{+0.36}$ &
$0.000_{-0.038}^{+0.050}$ &
$(1-10^{-6})_{-0.59}^{+10^{-6}}$ \\[0.8ex]
C &  
$-5.5_{-0.8}^{+1.9}$ &
$3.30_{-0.39}^{+0.94}$ &
$14.32_{-0.14}^{+0.36}$ &
$0.71_{-0.35}^{+0.27}$ &
$0.031_{-0.018}^{+0.017}$ &
$0.88_{-0.30}^{+0.12}$ \\[0.8ex]

\hline

\end{tabular}

\medskip
The simulated data are shown in Fig.~\ref{fig:OB66005} and corresponds to
the true parameters $t_\rmn{f}^\star = 0$,
$t_\star^\perp = 6~\mbox{h}$, $m_\rmn{f}^\star = 16.3$,
$g_\rmn{f}^\star = 0.05$ (${\hat t} = 1~\mbox{h}$), 
${\hat \omega}_\rmn{f}^\star = -0.001~\mbox{h}^{-1}$, and $\Gamma = 0.5$.
The letter
A denotes the outside region ($\eta < \eta_0^\rmn{out}$), 
B denotes the central region ($\eta_0^\rmn{out} \leq \eta <
\eta_0^\rmn{in}$), and C
denotes the inside region ($\eta \geq \eta_0^\rmn{in}$), 
with region B subdivided into B$_1$ ($\eta_0^\rmn{out} \leq \eta
< \eta_\rmn{c}$) and B$_2$ ($\eta_\rmn{c} \leq \eta < \eta_0^\rmn{in}$),
where $\Delta G_\rmn{f}^\star$ vanishes at $\eta_0^\rmn{out} = 0.716$ and 
$\eta_0^\rmn{in} = 1.762$, and $\eta_\rmn{c} = (\eta_0^\rmn{out}+\eta_0^\rmn{in})/2$.
Uppper-case letters refer to the dense sampling 
($\Delta t = 6~\mbox{min}$), while corresponding
lower-case letters refer to the sparse sampling ($\Delta t = 30~\mbox{min}$). The quoted uncertainties refer to intervals enclosing a probability 
of 68.3\,\%.
\end{minipage}
\end{table*}

Without going into much detail, \citet{Rhie:LD} have stated that a reliable
measurement of the limb-darkening coefficient requires a reasonable coverage
of the lightcurve in each of the regions A, B, and C. Using the
dataset presented in the previous subsections, let us study the information
content of each of these regions for the two different realizable
sampling rates 
and the adopted photometric precision that can be expected from the
PLANET campaign.

Table~\ref{tab:simu} gives the parameters obtained by the minimization of
$\chi^2$ for the simulated data shown in Fig.~\ref{fig:OB66005} in selected regions.
The quoted parameter uncertainties $\delta \vec p$
refer to projections of the hypersurface
$\Delta \chi^2 = \chi^2({\vec p_\rmn{min} + \delta \vec p}) - \chi^2(\vec p_\rmn{min}) = 1$ 
onto the parameter
axes, where $\vec p_\rmn{min}$ denotes the parameters at the $\chi^2$ minimum
$\chi^2_\rmn{min} = \chi^2(\vec p_\rmn{min})$, which correspond to intervals containing
a probability of 68.3\,\%. Since the model is not linear in the parameters, upper and
lower bounds differ, and 95.4\,\%-intervals ($\Delta \chi^2 = 4$)
are not twice the size of 68.3\,\%-intervals.
Table~\ref{tab:simuchi2} shows the corresponding values of $\chi^2_\rmn{min}$, the
probability $P_{\chi^2} = P(\chi^2 \geq \chi^2_\rmn{min})$ indicating
the goodness-of-fit, and the $\chi^2$ excess $\Delta\chi^2_\rmn{true}
= \chi^2(\vec p_{\rm true})-\chi^2(\vec p_\rmn{min})$ of the true
parameters $\vec p_{\rm true}$ with respect to the best-fit parameters
$\vec p_\rmn{min}$. The values of $\Delta\chi^2_\rmn{true}$ have to be seen with
regard to $\chi^2_\rmn{true}$ as listed in Table~\ref{tab:datasetprop} for the different regions 
of the lightcurve and the average misestimation
of error bars indicated by $\chi^2_\rmn{true}/N$.

Densely sampled data (every 6~min) for all of the regions (ABC) provide a
precise measurement of the limb-darkening coefficient $\Gamma$ with an uncertainty
of $\sim\,2.5\,$\% and for 
a sparser sampling (every 30~min) over all regions (abc), the
uncertainty  still remains below $\sim\,$6~\%.
For the dense (sparse) sampling, the uncertainties on the time $t_\rmn{f}^\star$ marking the
end of the caustic exit and on the passage half-duration $t_\star^\perp$ are
$\sim\,$30~s  ($\sim\,$90~s), while the flux at $t_\rmn{f}^\star$ is determined within
$\sim 0.4\,$\% ($\sim 1.5\,$\%) and
the relative uncertainties on $g_\rmn{f}$ and $\hat \omega_\rmn{f}^\star$ 
turn out to be $\sim 0.5\,$\% ($\sim 2\,$\%) and 
$\sim 10\,$\% ($\sim 40\,$\%), respectively.

From data in the outside region A alone, $\Gamma$ is determined with an uncertainty of 
$\sim\,$8~\%. While the slope discontinuity at the end of the caustic exit and the data thereafter
provide reliable measurements of $t_\rmn{f}^\star$, $m_\rmn{f}^\star$, and $\hat \omega_\rmn{f}^\star$ with
uncertainties not that much in excess of those for the full dataset, the uncertainties 
in $g_\rmn{f}^\star$ and $t_\star^\perp$ are largely increased by factors of 
$\sim\,$20 or $\sim\,60$, respectively.
By combining the outside region A with the inside region C, the latter parameters are accurately determined
and the uncertainties on all parameters are close to those for the full dataset (ABC), the uncertainty in
$\Gamma$ being $\sim\,$4~\%, so that the central region B does not add much valuable information if 
data in both of the regions A and C are available.

Data taken in each of the inside region C or the central region B alone is insufficient to provide a 
reliable measurement of the limb-darkening coefficient, where the central region B makes the worse case.
However, already a combination of the inside region C with the adjoining half of the central region B$_2$
yields $\Gamma$ with an uncertainty of $\sim\,$10~\%, which improves to $\sim\,$6~\% if the full region B
is included. The gain in the parameter uncertainties of the other parameters by adding region B$_2$ to
AB$_1$ is rather moderate, and the uncertainty on the prediction of the caustic exit time $t_\rmn{f}^\star$ from data
in regions AB$_1$ is $\sim\,$30~min, while the uncertainty on the corresponding magnitude is $\sim\,$2~mag.
Compared to the case in which only 
data in region A are used,
the uncertainties on $t_\rmn{f}^\star$, $m_\rmn{f}^\star$,
$g_\rmn{f}^\star$ and $\hat \omega_\rmn{f}^\star$  
are at least an order of magnitude larger, while the 
uncertainty on $t_\star^\perp$ however is about 3 times smaller.

Although data in region AC provide equal or better constraints than
data in region AB or AB$_1$ for all of the parameters
$t_\rmn{f}^\star$, $t_\star^\perp$, $m_\rmn{f}^\star$, $g_\rmn{f}^\star$ 
and $\hat \omega_\rmn{f}^\star$,
region AB$_1$ provides a measurement on $\Gamma$ of similar quality as region AC, and region
AB even provides the same small uncertainty of $\sim 2.5\,$\% as the full dataset ABC, making data
in region C obsolete in this case. 
Adding region B$_1$ to A allows a reliable determination of $t_\star^\perp$, and significantly reduces
the uncertainty in $g_\rmn{f}^\star$, whereas the uncertainty in $t_\rmn{f}^\star$ shows little and
those in $m_\rmn{f}^\star$ and ${\hat \omega}_\rmn{f}$ show no significant improvement.  
The main gain from region B$_2$ is in reducing the uncertainty in $t_\star^\perp$, whereas
$m_\rmn{f}^\star$ and ${\hat \omega}_\rmn{f}$ as well as their uncertainties remain practically unchanged,
and the improvements on $t_\rmn{f}^\star$ and $g_\rmn{f}^\star$ are quite small to negligible.

With dense sampling of region A yielding an uncertainty of $\sim\,8\,$\% on $\Gamma$ and regions B or C
being unable to provide a reasonable measurement, none of the regions 
alone could yield a sufficient accuracy
with the sparse sampling.  
From the combinations of two of these regions, 
ab yields a marginal limb-darkening measurement with an uncertainty of $\sim\,20\,$\%,
while the uncertainty is $\sim\,15\,$\%
 for bc and $\sim\,10\,$\% for ac. Nevertheless, one sees that it is possible
 to obtain a measurement of the linear limb-darkening coefficient
$\Gamma$ already with a small 
number of data points with high photometric accuracy 
over the course of the caustic passage.

The comparison of fits with at most one of the regions A, B, or C being densely and the remaining
parts of the source sparsely sampled shows that only dense sampling of region A yields a reasonable
reduction of the uncertainty of all parameters, where the uncertainty of the limb-darkening coefficient $\Gamma$ 
of $\sim\,3\,$\%  is close to that obtained for the densely sampled full dataset. In contrast,
the improvements by increasing
the sampling of regions B or C are marginal, although the better sampling of region C
yields significant reductions
in the uncertainties of $t_\star^\perp$, $g_\rmn{f}^\star$ and $\hat \omega_\rmn{f}^\star$.

As for the dense sampling of the outside region A, combining sparsely sampled data
in this region with data in one of the other regions is highly valuable, 
where in contrast to the
dense sampling, the inside region (C or c) provides more additional information 
(in particular through direct measurements of $t_\star^\perp$ and $g_\rmn{f}^\star$) than
the central region (B or b) and coverage of both regions contributes to a significant reduction
of the uncertainty on the limb-darkening coefficient $\Gamma$.
For both samplings of the outside region, most of the additional information
contained in the 
other regions is already extracted with the sparse sampling.
In all cases, the sampling rate in region A strongly affects the accuracy 
to which all parameters can be determined.

With data only in the inside and the central region being observed, a denser sampling
of the inside region strongly reduces the parameter uncertainties while
a denser sampling of the central region has very little effect.

\begin{table}
\caption{$\chi^2$ test on models based on
 data in different regions and corresponding $\chi^2$ excess of
true parameters.}
\label{tab:simuchi2}
\begin{tabular}{@{}ccccc}
\hline
Region & $\chi^2_\rmn{min}$ & d.o.f. & $P_{\chi^2}$ & $\Delta \chi^2_\rmn{true}$  
 \\
\hline
ABC & 160.1 & 147 & 0.22 & 5.4  \\
aBC & 115.8 & 108 & 0.29 & 3.9 \\ 
Abc & 54.7 & 60 & 0.67  & 7.3 \\ 
aBc & 82.2 & 74 & 0.24  & 4.5\\ 
abC & 46.2 & 55 & 0.80 & 4.0 \\ 
abc & 12.0 & 21 & 0.94 & 4.2  \\ 
AB$_1$ & 80.6 & 75 & 0.31 & 5.4 \\
AB & 123.2 & 105 &  0.11 & 5.6 \\
Ab &  55.1 & 52 &  0.36 & 7.6 \\
aB &  78.3 & 66 & 0.14 & 4.7 \\
ab &  8.4 & 13 & 0.82 & 5.1 \\
AC & 83.4 & 86 & 0.56 & 4.8 \\
Ac &  48.8 & 52 & 0.60 & 5.4 \\
aC &  39.1 & 47 &  0.79 & 3.3 \\
ac &  5.6 & 13 & 0.96 & 2.8\\
B$_2$C & 77.9 & 66 & 0.15 & 1.6\\ 
BC & 111.9 & 97 & 0.14 & 2.2 \\
Bc &  77.4 & 63 & 0.11 & 2.6\\
bC &  41.7 & 44 & 0.57 & 2.8 \\
bc &  7.5 & 10 & 0.68 & 3.0 \\
A & 44.4 & 44 & 0.45 &  7.0 \\ 
B & 74.3 & 55 & 0.04 & 3.0 \\ 
C & 32.3 & 36 & 0.65 & 4.5 \\ 
\hline
\end{tabular}

\medskip
The probability
$P_{\chi^2} = P(\chi^2 \geq \chi^2_\rmn{min})$ provides a 
measure of the goodness-of-fit,
while $\Delta \chi^2_\rmn{true} = \chi^2(\vec p_\rmn{true}) -
\chi^2(\vec p_\rmn{min})$ gives the excess in $\chi^2$ of the true
parameter set $\vec p_\rmn{true}$ over the best-fit parameter set
$\vec p_\rmn{min}$.
\end{table}

\section{Inappropriate brightness profiles}
\label{sec:wrongprofile}

\begin{table*}
\begin{minipage}{176mm}
\caption{Best-fit model parameters for the simulated dataset with
fixed inappropriate limb-darkening coefficient.}
\label{tab:parwronggam}
\begin{tabular}{@{}cccccccc}
\hline
$\Gamma$ & Region & 
$t_\rmn{f}^\star$ [h]& 
$t_\star^\perp$ [h]& $m_\rmn{f}^\star$ &
$g_\rmn{f}^\star$ & $-{\hat \omega}_\rmn{f}^\star$ [h$^{-1}$] \\
\hline
$0.3$ & ABC & 
$-0.1215_{-0.0037}^{+0.0037}$ &
$5.8791_{-0.0051}^{+0.0051}$ &
$16.2950_{-0.0041}^{+0.0041}$ &
$0.05030_{-0.00026}^{+0.00026}$ &
$0.00100_{-0.00011}^{+0.00011}$\\[0.8ex]
$0.3$ & abc & 
$-0.138_{-0.011}^{+0.011}$ &
$5.870_{-0.019}^{+0.019}$ &
$16.286_{-0.013}^{+0.014}$ &
$0.0507_{-0.0010}^{+0.0010}$ &
$0.00111_{-0.00044}^{+0.00052}$\\[0.8ex]
$0.7$ & ABC &  
$0.1259_{-0.0045}^{+0.0046}$ &
$6.1239_{-0.0060}^{+0.0060}$ &
$16.3028_{-0.0041}^{+0.0041}$ &
$0.04992_{-0.00025}^{+0.00025}$ &
$0.00094_{-0.00011}^{+0.00011}$\\[0.8ex]
$0.7$ & abc & 
$0.128_{-0.012}^{+0.012}$ &
$6.124_{-0.017}^{+0.018}$ &
$16.292_{-0.012}^{+0.012}$ &
$0.05026_{-0.00082}^{+0.00080}$ &
$0.00115_{-0.00039}^{+0.00045}$\\[0.8ex]
\hline
\end{tabular}

\medskip
Models with $\Gamma = 0.3$
or $\Gamma = 0.7$ using
the full simulated dataset corresponding to dense sampling (ABC)
or the reduced dataset corresponding to sparse sampling (abc).
The quoted uncertainties correspond to 68.3\,\% confidence intervals.
\end{minipage}
\end{table*}

\begin{table*}
\begin{minipage}{176mm}
\caption{$\chi^2$- and run test for models with a fixed inappropriate
limb-darkening coefficient.}
\label{tab:runtest}
\begin{tabular}{@{}ccccccccccccc}
\hline
$\Gamma$ & Region & $\chi^2_\rmn{min}$ & d.o.f. & $P_{\chi^2}$ &
$N$ & $N_{+}$ & $N_{-}$ & $\mathcal{E}(n_\rmn{r})$ &
$\sigma(n_\rmn{r})$ & $n_\rmn{r}^\rmn{obs}$
& $\delta$ & $P_\rmn{r}$ \\
\hline
0.3 & ABC & 566.7 & 148 & $2.8 \times 10^{-120}$ & 
153 & 77 & 76 & 77.50 & 6.16 & 48 & 4.79 & $8.3 \times 10^{-7}$ \\ 
0.3 & abc & 79.3 & 22 & $8.0 \times 10^{-10}$ &
27 & 10  & 17 & 13.59 & 2.37 & 7 & 2.78 & $2.7 \times 10^{-3}$ \\ 
0.7 & ABC & 419.6 & 148 & $1.8 \times 10^{-62}$ &
153 & 81 & 72 & 77.24 & 6.14 & 51 & 4.27 & $9.8 \times 10^{-6}$ \\ 
0.7 & abc & 39.3 & 22 & 0.013 &
27 & 19 & 8 & 12.26 & 2.11 & 10 & 1.07 & 0.14 \\ 
\hline
\end{tabular}

\medskip
Minimization of $\chi^2$ yielded $\chi^2_\rmn{min}$ for the listed models, which
corresponds to a probability $P_{\chi^2} = P(\chi^2 \geq \chi^2_\rmn{min})$.
Run tests for the same models and $N$ data points within the specified region
revealed $N_{+}$ positive and $N_{-}$ negative residuals forming 
$n_\rmn{r}^{\rmn{obs}}$ runs, where $\mathcal{E}(n_\rmn{r})$ runs are expected
with a standard deviation $\sigma(n_\rmn{r})$. The discrepancy 
$\delta = [\mathcal{E}(n_\rmn{r})- n_\rmn{r}^{\rmn{obs}}]/\sigma(n_\rmn{r})$
corresponds to a probability 
$P_\rmn{r} = P(n_\rmn{r} \leq 
n_\rmn{r}^\rmn{obs})$.  
\end{minipage}
\end{table*}

\begin{figure}
\includegraphics[width=84mm]{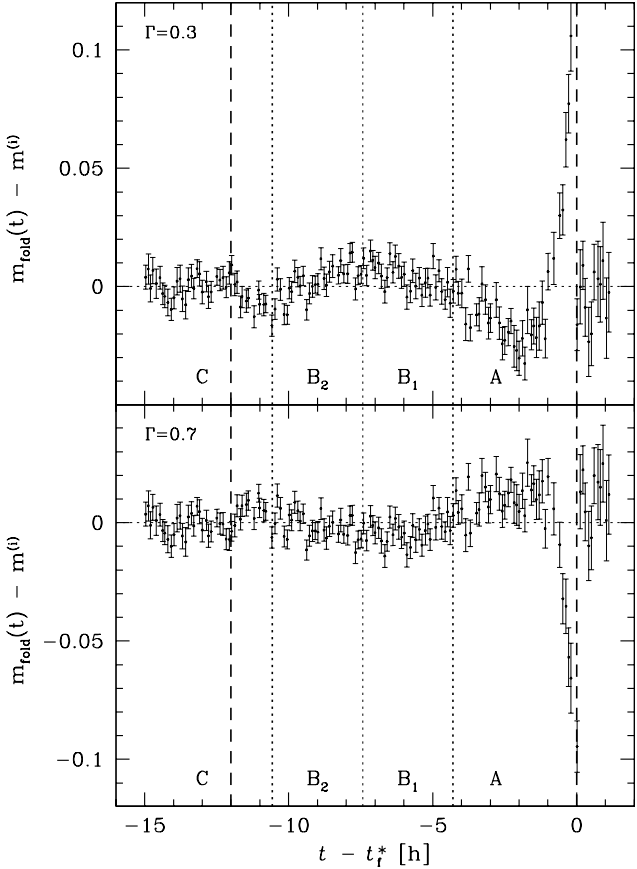}
\caption{Residuals of data with respect to model lightcurves for (incorrect) stellar brightness
profiles with $\Gamma = 0.3$ and $\Gamma = 0.7$, where the true value
of the underlying simulated dataset (as shown in Fig.~\ref{fig:OB66005}) is
$\Gamma = 0.5$.}
\label{fig:wronggam} 
\end{figure}

If an inappropriate brightness profile is adopted, the residuals of the data
with respect to the model lightcurve show characteristic systematic trends.
These are shown in 
Fig.~\ref{fig:wronggam} for models of the full simulated dataset 
displayed in Fig.~\ref{fig:OB66005} where the 
linear limb-darkening coefficient has been fixed at $\Gamma = 0.3$ or $\Gamma = 0.7$, 
respectively, whereas $\Gamma = 0.5$ marks its true value. The largest residuals
occur near the end of the caustic passage and there are long sequences
for which all data points either show larger or smaller magnifications than
the theoretical curve. In particular,
all data points near the time where the central
parts of the source passes deviate to one side, whereas points where
outer parts of the source pass deviate to the other side.
The sign of the residuals is just inverted for the 
two cases $\Gamma = 0.3$ and $\Gamma = 0.7$, representing a too weak or
a too strong linear limb-darkening term, respectively.

The parameters and their uncertainties for these two models and corresponding
models for the sparse sampling are shown
in Table~\ref{tab:parwronggam}. One sees that 
the achievement of the best-possible fit with the wrong brightness profile
leads to an inappropriate choice of the caustic passage half-duration $t_\star^\perp$
and the point of time $t_\rmn{f}^\star$ when the trailing limb of the source exits
the caustic, where $t_\star^\perp$ is underestimated
and $t_\rmn{f}^\star$ is shifted towards the caustic inside for
a weaker limb darkening ($\Gamma = 0.3$), whereas $t_\star^\perp$ is
overestimated and $t_\rmn{f}^\star$ is shifted towards the caustic outside
for a stronger limb darkening ($\Gamma = 0.7$). The remaining parameters
$m_\rmn{f}^\star$, $g_\rmn{f}^\star$, and ${\hat \omega}_\rmn{f}^\star$ are
fairly reproduced.

There is no significant difference in the residuals between either of the
two adopted samplings being applied 
to the different regions of data.
The same systematic trends still persist 
even if only data in the outside region A are modelled, although the size of 
the residuals is decreased compared to modelling the full dataset.

Table~\ref{tab:runtest} shows the results of both $\chi^2$- and run tests
for the models with fixed limb-darkening coefficient of the data
corresponding to the two different samplings.
The larger number of data points for the dense sampling strongly
increases the significance and each of the tests 
clearly recommends rejection of both of the models with $\Gamma = 0.3$
and $\Gamma = 0.7$. 
For the sparse sampling, this 
the case only for the model with $\Gamma=0.3$, whereas the model with
$\Gamma=0.7$ might be accepted. 
The $\chi^2$ tests turn out to be more powerful
than the run tests for all models. 
The power of the latter is limited by the requirement 
that subsequent systematic deviations to the same side need to be established
for each of the regions over which they persist separately, while the use
of the sum of the squared deviations in $\chi^2$ tests exploits the power of the
total number of data points, which can become quite large
for $N \ga 100$. However, as pointed out in Sect.~\ref{sec:stattests}, the uncertainty in the size of the true error
bars limits the power of rejecting models by means of a $\chi^2$ test.

While stellar brightness profiles described by a linear limb-darkening law have been studied
in detail in this section, it is obvious that other profiles show a similar behaviour
if the strength of limb darkening is under- or overestimated.

\section{Acceleration effects}
\label{sec:acceleration}

\begin{figure}
\includegraphics[width=84mm]{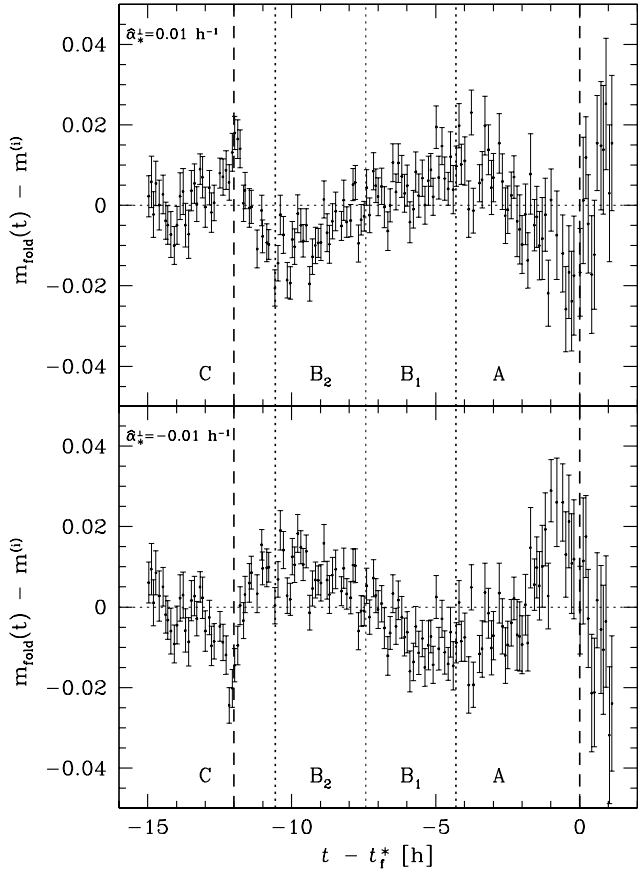}
\caption{Residuals of data with respect to model lightcurves with fixed
 (incorrect) acceleration
parameters 
${\hat \alpha}_{\star}^\perp = 0.01~\mbox{h}^{-1}$
and ${\hat \alpha}_{\star}^\perp = -0.01~\mbox{h}^{-1}$, where the true value
of the underlying simulated dataset (as shown in Fig.~\ref{fig:OB66005}) is
${\hat \alpha}_{\star}^\perp = 0$.} 
\label{fig:wrongacc}
\end{figure} 
\subsection{Origin and order estimates}

\begin{table*}
\begin{minipage}{176mm}
\caption{Estimates for the correlation coefficient between residuals
and elapsed time.}
\label{tab:corrcoff}
\begin{tabular}{@{}cccccccccc}
\hline
$\Gamma$ & ${\hat \alpha}_{\star}^\perp$ & 
$\hat \rho_1$ & $\hat \rho_2$ & $z_1$ & $z_2$ &
$\hat \rho^{+}$ & $\hat \rho^{-}$ & $z^{+}$ &
$z^{-}$  \\
\hline
0.3 & 0 & -0.59 & 0.62 & -0.61 & -0.68 & {\em 0.02} & -0.61 & {\em -0.03} & -0.70 \\
0.7 & 0 & 0.23 & -0.42 & 0.23 & -0.45 & {\em -0.10} & 0.32 & {\em -0.11} & 0.34 \\
0.5 & 0.01 &  0.38 & 0.55 & 0.40 & 0.62 & 0.46 & {\em -0.09} & 0.51 & {\em -0.11} \\ 
0.5 & 0.003 & {\em 0.03} & 0.24 & {\em 0.03} & 0.25 & 0.14 & {\em -0.11} &
  0.14 & {\em -0.11} \\
0.5 & 0.001 & {\em -0.08} & {\em -0.13} & {\em -0.08} & {\em -0.13} &
{\em 0.02} & {\em -0.11} & {\em 0.02} & {\em -0.11} \\ 
0.5 & -0.01 & -0.61 & -0.49 & -0.70 & -0.54 & -0.55 & {\em -0.06} &
 -0.62 & {\em -0.08} \\ 
0.5 & -0.003 & -0.32 & {\em -0.11} & -0.33 & {\em -0.11} &
-0.22 & {\em -0.10} & -0.22 & {\em -0.11} \\ 
0.5 & -0.001 & -0.20 & {\em 0.01} & -0.20 & {\em 0.01} &  {\em -0.10} & {\em -0.10} &
{\em -0.10} & {\em -0.11} \\ 
\hline
\end{tabular}

\medskip
For models involving the full dataset over regions A, B, and C, the
estimated correlation coefficients $\hat \rho_1$ and $\hat \rho_2$ correspond to data
in the subregions B$_1$ and B$_2$, while ${\hat \rho}^{\pm} = 
(\hat \rho_1\pm \hat \rho_2)/2$ give the 
antisymmetrized or symmetrized correlation for the full region
$B$. Symmetry is indicated by a vanishing $\hat \rho^{+}$, while a vanishing $\hat \rho^{-}$ indicates
antisymmetry.
With $N_1 = 31$ data points in the region B$_1$ and
$N_2 = 30$ data points in the region B$_2$, the standard deviation on 
 $z_{1,2} = 0.5\,\ln[(1+\hat \rho_{1,2})/(1-\hat \rho_{1,2}]$, is
$\sigma_{z_1} \approx \sigma_{z_1} \approx 0.19$ and the standard deviation on
$z^{\pm} = 
(z_1\pm z_2)/2$ is $\sigma_{z^{\pm}} \approx 0.13$. 
Values that are compatible with zero within one standard deviation are shown
in {\em italics}.
\end{minipage}
\end{table*}

The revolution of the Earth around the Sun and the orbital motion of the binary lens or a possible binary source \citep*{Do:Rotate,SMP2003} cause
an effective acceleration between lens and caustic, which leads to a modified 
relation between the caustic
passage phase $\eta$ and the elapsed time $t$, namely 
\begin{equation}
\eta = \pm\,\frac{t-t^\star_\rmn{f}}{t_\star^\perp}\,
\frac{1}{1+{\hat \alpha}_\star^\perp\,{t_\star^\perp}}
\,\left[
1 \pm \frac{{\hat \alpha}_\star^\perp}{2}\,
\left(t-t^\star_\rmn{f}\right)
\right]\,.
\end{equation}
This definition preserves the caustic passage taking place for $0 \leq \eta \leq 2$
and lasting $2\,t_\star^\perp$. The proper motion
of the source perpendicular to the caustic reads 
\begin{equation}
\mu^\perp(t) = \mu_\rmn{f}^\star\,\left[1+{\hat \alpha}_\star^\perp\,\left(t-t^\star_\rmn{f}
\right)\right]\,,
\end{equation}
where $\mu_\rmn{f}^\star = \mu^\perp(t^\star_\rmn{f})$ and with the constant acceleration
${\dot \mu}^\perp$, the acceleration parameter ${\hat \alpha}_\star^\perp$ is defined as
${\hat \alpha}_\star^\perp = {\dot \mu}^\perp/\mu_\rmn{f}^\star$.
For ${\hat \alpha}_\star^\perp < 0$, the source will reverse its direction of motion for
sufficiently large t where $\mu^\perp(t)$ becomes zero. Requiring that the source does
not move in opposite directions at $t^\star_\rmn{f} \pm 2\,t_\star^\perp$ and
$t^\star_\rmn{f}$ implies the restriction ${\hat \alpha}_\star^\perp \geq
-1/(2 t_\star^\perp)$ and ensures that $t^\star_\rmn{f} \pm 2\,t_\star^\perp$ is the
earlier time for which $\eta = 2$ is reached.

For microlensing observations toward the Galactic bulge, let us adopt source and lens 
distances of $D_\rmn{S} \sim 8.5~\mbox{kpc}$ and $D_\rmn{L} \sim 6~\mbox{kpc}$, 
respectively. With the Earth's velocity of $v_{\earth} \sim 30~\mbox{km}\,\mbox{s}^{-1}$, 
a change of the 
relative proper motion perpendicular to the caustic of ${\dot \mu}^\perp_{\earth}
= \kappa_{\earth}^\perp\,\frac{D_\rmn{S} - D_\rmn{L}}{D_\rmn{L}\,D_\rmn{S}}\,
(v_{\earth}^2/(1~\mbox{AU})) \sim 1.5 \times 10^{-2}~\mbox{$\mu$as}\,\mbox{d}^{-1}$ is induced, where
$\kappa_{\earth}^{\perp} \leq 1$ is a geometrical projection factor depending on the angles between
the Earth's acceleration vector and the line-of-sight and between its projection perpendicular to
the line-of-sight and the relative proper motion between source and caustic.
For a relative perpendicular proper motion of $\mu^\perp \sim 10~\mbox{$\mu$as}\,\mbox{d}^{-1}$,
corresponding to a velocity of $v^\perp = D_\rmn{L}\,\mu^\perp \sim 100~\mbox{km}\,\mbox{s}^{-1}$ at the position of the lens,
the acceleration parameter becomes ${\hat \alpha}_{\star,\earth}^\perp = 
{\dot \mu}^\perp_{\earth}/\mu^\perp \sim 4 \times 10^{-5}\,\kappa_{\earth}^{\perp}~\mbox{h}^{-1}$.

The orbital motion of a binary lens causes
a rotation of its caustic with its orbital period and an alteration of shape and
strength due to changes in the angular separation between its constituents.
For a total
mass $M \sim 0.7~M_{\sun}$ and semi-major axis $a \sim 1.6~\mbox{AU}$, the orbital period
becomes 
$P \sim 2.4~\mbox{yr}$, the  
Einstein radius for $D_\rmn{S} \sim 8.5~\mbox{kpc}$ and $D_\rmn{L} \sim 6~\mbox{kpc}$
is $r_\rmn{E} \sim 3.1~\mbox{AU}$, and the angular Einstein radius
is $\theta_\rmn{E} \sim 0.54~\mbox{mas}$.
For moderate eccentricities, the orbital acceleration does not differ much from that for
circular orbits. Depending on the orbital inclination and phase, the separation between
the constituents of the binary lens falls into the range $0 \leq r_\rmn{p} \leq a$, where
$r_\rmn{p} = a$ for all times in face-on orbits. 
Since $a \sim 0.51~r_\rmn{E}$,
the lens is a close binary with respect to its
caustic topology for any mass ratio $q$, 
showing a diamond-shaped caustic at its center-of-mass \citep{Erdl,Do99:CR}.
The acceleration induced by the rotation for a point on the caustic located at
an angle $\theta$ from the center-of-mass reads
${\dot \mu}^\perp_{\rmn{rot}}
= 4 \upi^2\,\kappa^\perp_{\rmn{rot}}\,(r_\rmn{p}/a)\,P^{-2}\,\theta$,
where the factor $r_\rmn{p}/a$ accounts for the projection of the orbital motion onto the sky,
while the geometrical factor $\kappa^\perp_{\rmn{rot}} \leq 1$ is related to the angle between the projected
acceleration vector and the proper motion between source and caustic.
For a mass ratio $q \sim 0.8$ and $r_\rmn{p} = a$, the angular extent of the caustic is $\sim 0.26\,
\theta_\rmn{E}$ \citep{Do99:CR}, and for a
source intersecting the caustic at $\theta \sim 0.1~\theta_\rmn{E}$,
the angular acceleration perpendicular to the caustic becomes
${\dot \mu}^\perp_{\rmn{rot}}
= 2.7 \times 10^{-3}\,\kappa^\perp_{\rmn{rot}}~\mbox{$\mu$as}\,\mbox{d}^{-1}$, so that
$\mu^\perp \sim 10~\mbox{$\mu$as}\,\mbox{d}^{-1}$
implies an acceleration parameter of 
${\hat \alpha}_{\star,\rmn{rot}}^\perp  
 \sim 10^{-5}\,\kappa^{\perp}_{\rmn{rot}}~\mbox{h}^{-1}$.

While face-on orbits provide the largest effect on the acceleration from the rotation, 
the alteration of the angular separation of the constituents is largest for edge-on orbits and
causes the size of the caustic to change proportional to $d^2 = (r_{\rm p}/r_E)^2$.
With $(\rmn{d}^2/\rmn{d}t^2) d^2 = 2 (d\ddot{d} + {\dot d}^2)$ and $d(t) = (a/r_{\rm E})\,
\cos[(2\pi/P)t+\varphi_0]$ for a circular edge-on orbit,
the induced acceleration for a point on the caustic at angle $\theta$ reads
${\dot \mu}^\perp_{\rmn{osc}}
= 8 \upi^2\,\kappa^\perp_{\rmn{osc}}\,P^{-2}\,\theta$, which becomes
${\dot \mu}^\perp_{\rmn{osc}}
= 5.4 \times 10^{-3}\,\kappa^\perp_{\rmn{osc}}~\mbox{$\mu$as}\,\mbox{d}^{-1}$ for the parameters
chosen above, so that the acceleration parameters becomes
${\hat \alpha}_{\star,\rmn{osc}}^\perp  
\sim  2\times10^{-5}\,\kappa^{\perp}_{\rmn{osc}}~\mbox{h}^{-1}$, where the 
 geometrical factor $\kappa^{\perp}_{\rmn{osc}}\leq 1$ accounts for the orbital phase and 
inclination as well as for the caustic crossing angle.

The effective acceleration caused by orbital motion strongly 
increases towards shorter binary periods,
however, in order to intersect the increasingly smaller caustics, 
smaller impact parameters are required, which are less likely to occur. Larger acceleration
effects are therefore not impossible but rather improbable.
In general, the contributions of parallax and the rotational and oscillatory contributions
by orbital motion are rougly of competitive order.

\subsection{Signature of acceleration}

\begin{table*}
\begin{minipage}{176mm}
\caption{Best-fit model parameters obtained for selected fixed acceleration parameters.}
\label{tab:parwrongacc}
\begin{tabular}{@{}cccccccc}
\hline
${\hat \alpha}_{\star}^\perp$ [h$^{-1}$]& Region & 
$t_\rmn{f}^\star$ [h]& 
$t_\star^\perp$ [h]& $m_\rmn{f}^\star$ &
$g_\rmn{f}^\star$ & $-{\hat \omega}_\rmn{f}^\star$ [h$^{-1}$] & $\Gamma$ \\
\hline
$0.01$ & ABC &  
$-0.0044_{-0.0086}^{+0.0080}$ &
$5.9103_{-0.0087}^{+0.0090}$ &
$16.2916_{-0.0039}^{+0.0039}$ &
$0.05073_{-0.00024}^{+0.00024}$ &
$0.000228_{-0.000082}^{+0.000084}$ &
$0.423_{-0.011}^{+0.011}$\\[0.8ex]
$0.003$ & ABC &  
$-0.0005_{-0.0089}^{+0.0092}$ &
$5.9731_{-0.0096}^{+0.0098}$ &
$16.2951_{-0.0040}^{+0.0040}$ &
$0.05037_{-0.00025}^{+0.00024}$ &
$0.00073_{-0.00010}^{+0.00010}$ &
$0.485_{-0.012}^{+0.012}$\\[0.8ex]
$0.001$ & ABC &  
$0.0003_{-0.0082}^{+0.0088}$ &
$5.9919_{-0.0083}^{+0.0091}$ &
$16.2962_{-0.0040}^{+0.0040}$ &
$0.05026_{-0.00024}^{+0.00024}$ &
$0.00090_{-0.00010}^{+0.00011}$ &
$0.503_{-0.012}^{+0.012}$\\[0.8ex]
$-0.01$ & ABC & 
$0.013_{-0.010}^{+0.010}$ &
$6.123_{-0.012}^{+0.011}$ &
$16.3076_{-0.0044}^{+0.0045}$ &
$0.04903_{-0.00032}^{+0.00031}$ &
$0.00248_{-0.00021}^{+0.00022}$ &
$0.612_{-0.014}^{+0.013}$\\[0.8ex]
$-0.003$ & ABC &  
$-0.0009_{-0.0080}^{+0.0096}$ &
$6.029_{-0.008}^{+0.011}$ &
$16.2985_{-0.0040}^{+0.0040}$ &
$0.05002_{-0.00025}^{+0.00024}$ &
$0.00013_{-0.00011}^{+0.00013}$ &
$0.536_{-0.012}^{+0.013}$\\[0.8ex]
$-0.001$ & ABC &  
$0.0004_{-0.0089}^{+0.0092}$ &
$6.0108_{-0.0098}^{+0.0099}$ &
$16.2973_{-0.0040}^{+0.0040}$ &
$0.05014_{-0.00025}^{+0.00025}$ &
$0.00108_{-0.00011}^{+0.00012}$ &
$0.521_{-0.012}^{+0.012}$\\[0.8ex]
\hline
\end{tabular}

\medskip
Best-fit parameters and 68.3\,\% confidence intervals
for models involving selected fixed values for the acceleration parameter 
${\hat \alpha}_{\star}^\perp$
using the full dataset corresponding to dense sampling for which 
${\hat \alpha}_{\star}^\perp$ had been adopted.
\end{minipage}
\end{table*}

\begin{table*}
\begin{minipage}{176mm}
\caption{$\chi^2$- and run test for models 
with fixed acceleration parameters.}
\label{tab:runtestacc}
\begin{tabular}{@{}ccccccccccccc}
\hline
${\hat \alpha}_{\star}^\perp$ [h$^{-1}$]& Region & $\chi^2_\rmn{min}$ & d.o.f. & $P_{\chi^2}$ &
$N$ & $N_{+}$ & $N_{-}$ & $\mathcal{E}(n_\rmn{r})$ &
$\sigma(n_\rmn{r})$ & $n_\rmn{r}^\rmn{obs}$
& $\delta$ & $P_\rmn{r}$ \\
\hline
0.01 & ABC &  439.0 & 147 & $4.3 \times 10^{-70}$ & 153 & 78 & 75 & 77.47 & 6.16 & 39 & 6.24 & $2.1 \times 10^{-10}$ \\ 
0.003 & ABC & 191.6 & 147 & $7.8 \times 10^{-3}$ & 153 & 77 & 76 & 77.49 & 6.16 & 75 & 0.41 & 0.34 \\ 
0.001 & ABC & 164.5 & 147 & 0.15 & 153 & 73 & 80 & 77.33 & 6.15 & 83 & -0.92 & 0.82 \\ 
-0.01 & ABC & 496.7 & 147 & $3.3 \times 10^{-92}$ & 153 & 78 & 75 & 77.47 & 6.16 & 44 & 5.43 & $2.8 \times 10^{-8}$ \\ 
-0.003 & ABC & 185.2 & 147 & 0.018 & 153 & 75 & 78 & 77.47 & 6.16 & 70 & 1.21 & 0.11 \\ 
-0.001 & ABC & 161.9 & 147 & 0.19 & 153 & 77 & 76 & 77.50 & 6.16 & 73 & 0.73 & 0.23 \\ 
\hline
\end{tabular}

\medskip
On the full set of $N =153$ data points, the $\chi^2$-minimization yielded $\chi^2_\rmn{min}$ with
$P_{\chi^2} = P(\chi^2 \geq \chi^2_\rmn{min})$ and the run test revealed 
$N_{+}$ positive and $N_{-}$ negative residuals, for which $\mathcal{E}(n_\rmn{r})$ runs
with a standard deviation $\sigma(n_\rmn{r})$ are expected, whereas  
$n_\rmn{r}^{\rmn{obs}}$ have been found, yielding 
$\delta = [\mathcal{E}(n_\rmn{r})-n_\rmn{r}^{\rmn{obs}}]/
\sigma(n_\rmn{r})$, corresponding to a probability $P_\rmn{r} = P(n_\rmn{r} \leq 
n_\rmn{r}^\rmn{obs})$.  
\end{minipage}
\end{table*}

In order to cause deviations of 1--2\,\%, the acceleration parameter
needs to exceed $|{\hat \alpha}_{\star}^\perp| \ga 0.01~\mbox{h}^{-1}$
for the parameters $g_\rmn{f}^\star = 0.05$ (${\hat t} = 1~\mbox{h}$) and $t_\rmn{f}^\star =
6~\mbox{h}$, which have been chosen for the simulated dataset,
and is therefore required
to be much larger than the expected values, indicating that effective
acceleration is unlikely to play an important role.
The influence of acceleration increases with
\begin{eqnarray}
f_\rmn{acc} & \propto & {\hat \alpha}_{\star}^\perp\,
\left({t_\star^\perp}\right)^{1/2}Ê\left(g_\rmn{f}^\star\right)^{-1} \nonumber \\
& \propto & {\hat \alpha}_{\star}^\perp\,t_\star^\perp\,\theta_\star^{-1/2}\,
10^{-0.4(m_\rmn{S} - m_\rmn{f}^\star)} \nonumber \\
& = & {\hat \alpha}_{\star}^\perp\,
\left(t_\star^\perp/\mu^\perp\right)^{1/2}\,
10^{-0.4(m_\rmn{S} - m_\rmn{f}^\star)}\,,
\end{eqnarray}
so that stronger effects occur for longer passage half-durations $t_\star^\perp$ 
in combination with smaller angular stellar radii $\theta_\star$ 
or smaller proper motions $\mu^\perp$, as well as for larger caustic strength $R_\rmn{f}$,
although a very large increase is not likely to occur, whereas a decrease results from
larger $m_\rmn{S} - m_\rmn{f}^\star$, i.e.\ larger magnification at $t_\rmn{f}^\star$.

The residuals for models with 
${\hat \alpha}_{\star}^\perp = \pm 0.01~\mbox{h}^{-1}$, as shown
in Fig.~\ref{fig:wrongacc}, exhibit characteristic systematic trends,
which can be distinguished from those occuring with the adoption of an
inappropriate stellar brightness profile.
For the latter case, the residuals during the central
part of the caustic passage (region B) are roughly symmetric with respect to the
boundary between the subregions B$_1$ and B$_2$
while deviations to one side dominate for
adjoining parts of the limb regions A and C,
and just the very outer limb in region A shows a strong deviation 
to the same side as the central part of region B.
In contrast, the residuals for models including an inappropriate assumed
acceleration 
are roughly antisymmetric about the temporal midpoint
for each of the regions A, B, and C, showing a linear trend and reaching
maxima at about the boundaries of these regions and the endpoints of the
caustic passage.
Therefore, tests of symmetry or antisymmetry in the different
regions can be used to decide whether model discrepancies are due to
acceleration effects or the adoption of inappropriate brightness profiles.

For $N$ data points $(x_i, y_i)$, where $1 \leq i \leq N$, an estimate for
the correlation coefficient $-1 \leq \hat\rho \leq 1$ is obtained as
\begin{equation}
\hat \rho = \frac{N \sum x_i y_i - \sum x_i \sum y_i}{
\sqrt{\left[N \sum x_i^2 - (\sum x_i)^2\right]\left[N \sum y_i^2 - (\sum y_i)^2\right]}}\,,
\end{equation}
where Fisher's transform 
\begin{equation}
z = \frac{1}{2}\,\ln\frac{1+\hat \rho}{1-\hat \rho}
\end{equation}
asymptotically follows a normal distribution around its expectation value with
standard deviation $\sigma_z = 1/\sqrt{N-3}$.

For the subregions
B$_1$ and B$_2$, estimates for the correlation coefficient,
$\hat \rho_1$ and
$\hat \rho_2$, and the corresponding values of Fisher's transform,
$z_1$ and $z_2$, have been calculated 
for the previously discussed models involving an incorrect brightness profile
or acceleration parameter based on the complete dataset ABC. The antisymmetrized
or symmetrized correlation over the whole region B is characterized by
$\hat \rho^{\pm} = (\hat \rho_1 \pm \hat \rho_2)/2$ 
and $z^{\pm} = (z_1 \pm z_2)/2$ with
$\sigma_{z^{\pm}} = 0.5\,\sqrt{(N_1+N_2-6)/[(N_1-3)(N_2-3)]}$.
For a symmetry (as for
an incorrect brightness profile), one expects $\hat \rho^{+}$ 
to vanish, while
$\hat \rho^{-}$ is expected to vanish for an antisymmetry 
(as for an incorrect acceleration parameter), whereas signatures of a symmetry or
antisymmetry are obtained from values that significantly differ from zero.
The obtained values are displayed in Table~\ref{tab:corrcoff}, where
entries in {\em italics}{} are compatible with zero within the standard deviation.
The obtained estimates for the correlation coefficient
reflect the expected behaviour and allow to distinguish 
between the two different causes of systematic residuals unless the random scatter
of the residuals dominates over the systematic trend, which is the case for 
$|{\hat \alpha}_\star^\perp| = 0.001~\mbox{h}^{-1}$, where both 
$\hat \rho^{+}$ and $\hat \rho^{-}$ are compatible with zero.

The model parameters and their 68.3\,\% confidence intervals are shown in
Table~\ref{tab:parwrongacc}. One sees that larger acceleration parameters yield
smaller limb-darkening coefficients $\Gamma$, shorter caustic passage
half-durations $t_\star^\perp$, smaller ${\hat \omega}_\rmn{f}^\star$ and
larger $g_\rmn{f}^\star$, whereas smaller acceleration
parameters alter the estimates for $\Gamma$, $t_\star^\perp $, 
${\hat \omega}_\rmn{f}^\star$, and $g_\rmn{f}^\star$ in the opposite
direction.

While probabilities corresponding to $\chi^2$ as shown in Table~\ref{tab:runtestacc}
reveal the incompatibility of the model with the data for the selected values
of ${\hat \alpha}_\star^\perp = \pm 0.01~\mbox{h}^{-1}$ and 
${\hat \alpha}_\star^\perp = \pm 0.003~\mbox{h}^{-1}$ due to the size of the
absolute deviations, the results of a run test show the presence of systematic
trends for ${\hat \alpha}_\star^\perp = \pm 0.01~\mbox{h}^{-1}$. 
As can be seen from both Table~\ref{tab:parwrongacc} and Table~\ref{tab:runtestacc}
and the comparison with the models parameters and their uncertainties
for vanishing acceleration as listed in Table~\ref{tab:simu},
the effects of acceleration
become negligible for ${\hat \alpha}_\star^\perp = \pm 0.001~\mbox{h}^{-1}$.

\subsection{Influence on limb-darkening measurement}

\begin{table*}
\begin{minipage}{176mm}
\caption{Best-fit model parameters including acceleration and their uncertainties
for the simulated data
in selected regions.}
\label{tab:paracc}
\begin{tabular}{@{}cccccccc}
\hline
Region &  $t_\rmn{f}^\star$ [h]& 
$t_\star^\perp$ [h]& $m_\rmn{f}^\star$ &
$g_\rmn{f}^\star$ & $-{\hat \omega}_\rmn{f}^\star$ [h$^{-1}$] & 
${\hat \alpha}_{\star}^\perp$ [h$^{-1}$] & $\Gamma$ \\
\hline
ABC & 
$0.0000_{-0.0089}^{+0.0091}$ &
$6.003_{-0.011}^{+0.011}$ &
$16.2968_{-0.0040}^{+0.0040}$ &
$0.05019_{-0.00025}^{+0.00025}$ &
$0.00100_{-0.00012}^{+0.00013}$ &
$-0.00020_{-0.00057}^{+0.00057}$ &
$0.513_{-0.013}^{+0.013}$ \\[0.8ex]
abc & 
$-0.001_{-0.022}^{+0.023}$ &
$6.014_{-0.026}^{+0.023}$ &
$16.292_{-0.013}^{+0.013}$ &
$0.05012_{-0.00085}^{+0.00085}$ &
$0.00140_{-0.00042}^{+0.00046}$ &
$-0.0009_{-0.0011}^{+0.0012}$ &
$0.526_{-0.031}^{+0.031}$ \\[0.8ex]
ab &
$0.06_{-0.16}^{+0.11}$ &
$5.96_{-0.13}^{+0.13}$ &
$16.41_{-0.22}^{+0.17}$ &
$0.0426_{-0.006}^{+0.027}$ &
$0.011_{-0.018}^{+0.010}$ &
$0.0011_{-0.0056}^{+0.0079}$ &
$0.44_{-0.14}^{+0.10}$ \\[0.8ex]
ac &
$-0.017_{-0.026}^{+0.028}$ &
$6.002_{-0.027}^{+0.025}$ &
$16.291_{-0.013}^{+0.013}$ &
$0.05026_{-0.00088}^{+0.00088}$ &
$0.00133_{-0.00043}^{+0.00047}$ &
$-0.0008_{-0.0014}^{+0.0015}$ &
$0.494_{-0.042}^{+0.043}$ \\[0.8ex]
BC & 
$0.0_{-1.2}^{+0.5}$ &
$6.00_{-0.60}^{+0.26}$ &
$16.4_{-1.4}^{+2.4}$ &
$0.04_{-0.04}^{+0.14}$ &
$0.0008_{-0.0013}^{+0.0093}$ &
$-0.0017_{-0.0080}^{+0.0060}$ &
$0.506_{-0.069}^{+0.064}$ \\[0.8ex]
A & 
$-0.026_{-0.021}^{+0.026}$ &
$4.5_{-0.9}^{+4.9}$ &
$16.3008_{-0.0070}^{+0.0070}$ &
$0.066_{-0.035}^{+0.017}$ &
$0.00173_{-0.00095}^{+0.00092}$ &
$0.073_{-0.084}^{+0.014}$ &
$0.466_{-0.043}^{+0.044}$ \\[0.8ex]
\hline

\end{tabular}

\medskip
The true parameters are $t_\rmn{f}^\star = 0$,
$t_\star^\perp = 6~\mbox{h}$, $m_\rmn{f}^\star = 16.3$,
$g_\rmn{f}^\star = 0.05$ (${\hat t} = 1~\mbox{h}$), 
${\hat \omega}_\rmn{f}^\star = -0.001~\mbox{h}^{-1}$, 
${\hat \alpha}_{\star}^\perp = 0$ and $\Gamma = 0.5$.
The quoted uncertainties refer to intervals enclosing a probability 
of 68.3\,\%.
\end{minipage}
\end{table*}

In general, acceleration can be assessed by including ${\hat \alpha}_{\star}^\perp$ as additional
free parameter in models of the lightcurve. If it cannot be accurately determined, degeneracies with
other model parameters can lead to an increase of their uncertainties. 
Although variations of the amount of limb darkening and
of acceleration provide different signatures over the full course of the caustic passage as
previously shown, similar effects in local regions of the lightcurve can arise so that
with a partial sampling
their origin cannot be resolved and the measurement of the limb-darkening coefficient is blurred
by the possible presence of significant acceleration.  

The uncertainties on best-fit model parameters including a free acceleration parameter
${\hat \alpha}_{\star}^\perp$
based on the simulated dataset for several selected regions of the lightcurve
shown in Table~\ref{tab:paracc} and the results of a 
$\chi^2$ test shown in Table~\ref{tab:chi2acc} demonstrate that
acceleration affects models for which the caustic passage duration cannot be
determined with sufficient accuracy, 
while otherwise a precise measurement of the acceleration leaves both the other model
parameters and their uncertainties practically unchanged and the
uncertainty of the measurement of the linear limb-darkening coefficient does
not suffer from acceleration effects as a comparison with models for the same regions
that neglect 
acceleration effects listed in Table~\ref{tab:simu} shows.

Even with the adopted dense sampling of the outside region A alone, 
for which the acceleration parameter is compatible with values of the order
of $|{\hat \alpha}_{\star}^\perp| \sim 0.1~\mbox{h}^{-1}$,
a meaningful measurement
of $\Gamma$ cannot be obtained and large parameter degeneracies arise
if acceleration is considered to be a free parameter.
In contrast, if data in the outside region are combined with data
from at least one of the other regions,
the amount of acceleration is accurately determined and therefore does not influence the uncertainties
on other parameters already for the more sparse sampling. 
Without a dense sampling of the outside region, 
the remaining parts of the
lightcurve need to be densely sampled in order to yield 
a meaningful measurement of the linear
limb-darkening coefficient, 
where its uncertainty is increased by a factor of two
compared to the assumption of vanishing acceleration.

Data taken over all regions allows to assess the acceleration parameter accurately,
down to $|{\hat \alpha}_{\star}^\perp| \la 10^{-3}~\mbox{h}^{-1}$ for the sparse
sampling and even down to $|{\hat \alpha}_{\star}^\perp| \la 6 \times 10^{-4}~\mbox{h}^{-1}$ 
for the dense sampling, which is sufficient for acceleration effects
effects on the lightcurve and the determination of the other parameters and their
uncertainties to become negligible.

In any case, parameter degeneracies involving acceleration can be limited by adopting an uppper limit
on $|{\hat \alpha}_{\star}^\perp|$ resulting from an assessment of the dynamics of the binary lens
and the Earth's motion at the time of observations.

\begin{table}
\caption{$\chi^2$ test on models including acceleration for simulated data in
selected regions and corresponding $\chi^2$ excess of
true parameters.}
\label{tab:chi2acc}
\begin{tabular}{@{}ccccc}
\hline
Region & $\chi^2_\rmn{min}$ & d.o.f. & $P_{\chi^2}$ & $\Delta \chi^2_\rmn{true}$  
 \\
\hline
ABC & 160.0 & 146 & 0.20 & 5.6  \\
abc & 11.4 & 20 & 0.93 & 4.8  \\ 
ab &  8.4 & 12 & 0.75 & 5.1 \\
ac &  5.3 & 12 & 0.95 & 3.1\\
BC & 111.8 & 96 & 0.13 & 2.3 \\
A & 43.8 & 43 & 0.44 &  7.7 \\ 
\hline
\end{tabular}

\medskip
The probability
$P_{\chi^2} = P(\chi^2 \geq \chi^2_\rmn{min})$ provides a 
measure of the goodness-of-fit,
while $\Delta \chi^2_\rmn{true} = \chi^2(\vec p_\rmn{true}) -
\chi^2(\vec p_\rmn{min})$ gives the excess in $\chi^2$ of the true
parameter set $\vec p_\rmn{true}$ over the best-fit parameter set
$\vec p_\rmn{min}$.
\end{table}

\section{Summary and recommendations on observing strategy}
\label{sec:summary}

As laid out by \citet{SW1987}, the differential magnification across the face of a
source during its passage over a caustic created by an intervening gravitational lens
can reveal its radial brightness profile through frequent accurate photometric observations.
Galactic microlensing surveys such as OGLE-III are capable of providing
 $\sim\,10$ such passages of stars in the Galactic bulge per
year from which a measurement of the stellar brightness profile can be obtained, which 
reflects the variation of temperature
with distance from the center providing a powerful
technique for probing stellar atmosphere models.

Contrary to caustic entries, the corresponding exits are predictable from data obtained
during the characteristic rise 
in magnification to a peak \citep{PLANET:sol,Mao:exit}, which allows to schedule 
frequent observations before the caustic passage begins.

While the caustic passages are likely to last from a few hours to a few days,
photometric measurements with 
1m-class telescopes with an uncertainty of less than 1.5~\% with
sampling intervals of a few minutes can be obtained for stars brighter than $\sim\,17$th
magnitude as currently being carried out by the PLANET collaboration \citep{PLANET:EGS},
offering the possibility to collect a few hundred data points during 
the course of the passage.

In the vicinity of the caustic passage, the lightcurve can be described by means of a 
characteristic profile function $G_\rmn{f}^\star(\eta;\xi)$, which depends solely
on the dimensionless normalized stellar brightness profile $\xi(\rho)$ as function
of the fractional radius $\rho$ and the caustic passage phase $\eta$. 
$G_\rmn{f}^\star(\eta;\xi)$ can be seen as the response delivered by the caustic
to the specific form of the brightness profile. The weight of the contribution of the
brightness at a specific fractional radius $\rho$ to the caustic profile function
$G_\rmn{f}^\star(\eta; \xi)$ is given by a function ${\bmath {\mathcal T}}(\eta, \rho)$,
which measures the sensitivity of the lightcurve at the point of time that corresponds
to the caustic passage phase $\eta$ to a local variation of the stellar brightness 
profile at fractional radius $\rho$. An inspection of ${\bmath {\mathcal T}}(\eta, \rho)$ shows
that the caustic passage provides a one-dimensional scan of the brightness profile where
each fractional radius is most efficiently probed as it touches the caustic, which
happens twice during the course of the caustic passage, so that the
majority of information about outer radii is provided at the beginning and the end of the caustic passage,
while inner regions of the source reveal their identity close to times when the source center
passes. However, the integrated sensitivity over the full duration of the
caustic passage increases with
fractional radius.

In general, the extraction of the stellar brightness profile  $\xi(\rho)$ from the observed lightcurve
$m_\rmn{fold}(t)$ involves the determination of 5 model parameters 
($t_\rmn{f}^\star$, $t_\star^\perp$, $m_\rmn{f}^\star$, $g_\rmn{f}^\star$, $\hat \omega_\rmn{f}^\star$),
which relate $m_\rmn{fold}(t)$ and the caustic profile function $G_\rmn{f}^\star(\eta;\xi)$, as well as
the solution of the integral equation, which relates $G_\rmn{f}^\star(\eta;\xi)$ and $\xi(\rho)$ by
means of the caustic sensitivity function ${\bmath {\mathcal T}}(\eta, \rho)$.

The variation of coefficients characterizing a parametrized stellar brightness profile causes 
variations in the observed lightcurve through variations of the brightness profile at all fractional radii.
The sensitivity of the lightcurve to variations in such a coefficient $\Gamma$ therefore depends
both on the
properties of ${\bmath {\mathcal T}}(\eta, \rho)$ and $\partial \xi(\rho; \Gamma)/\partial \Gamma$. 
For linear limb darkening, this sensitivity is
largest near the beginning of a caustic exit, followed by the surroundings of its end,
whereas the smaller integrated sensitivity over the caustic passage for smaller fractional radii and
the fact that $\partial \xi(\rho; \Gamma)/\partial \Gamma$ shows little variation with $\rho$ for the
inner parts of the source  
imply a small sensitivity while these pass the caustic.  
However, the identification of the end of the caustic exit with the characteristic feature of a jump discontinuity
in its slope allows a direct measurement of the corresponding point of time $t_\rmn{f}^\star$ and magnitude 
$m_\rmn{f}^\star$, while data points just after the end of the caustic exit can be used to measure
the parameter $\hat \omega_\rmn{f}^\star$ directly, which characterizes 
the variation of magnification due to non-critical images of the source. 
For this reason, sampling of the surroundings of the end of the caustic exit is more valuable than of its 
beginning. Monitoring of both of these regions provides direct measurements
of the caustic passage half-duration $t_\star^\perp$ and the caustic rise parameter $g_\rmn{f}^\star$.

Unless a precision measurement on the linear limb-darkening coefficient $\Gamma$ is desired, a sampling interval
at the limit of the capabilities of the monitoring campaign is not required. 
For a caustic passage lasting 12~h, a sampling interval of $\sim\,$30~min 
(corresponding to  $\sim\,$25--30 data points) is sufficient to provide 
$\Gamma$ with an uncertainty of less than $\sim\,$8~\%, while a dense sampling corresponding to
an interval of $\sim\,$6~min (with $\sim\,$150 data points)
would reduce the uncertainty to less than $\sim\,$3~\%.
This offers the possibility
for measuring limb-darkening coefficients in several broadband filters with the same telescope on the
same microlensing event or to monitor other microlensing events during the same night (e.g.\ to look for
anomalies caused by planets around the lens stars).

Since coverage of the surroundings of the end of a caustic exit is most effective in
providing an accurate measurement of the linear limb-darkening coefficient, 
it should be the goal of any observation strategy to try to obtain data in this region,
where the sampling rate should be chosen with regard to the desired accuracy of the measurement
and the number of broadband filters.
Moderate sampling over other regions of the light curve
is more valuable than an increased sampling over the end of the caustic exit, where a very dense
sampling over these other regions however
does not provide much additional information but can be used to compensate
for a missing coverage of some parts of the caustic passage.

The adoption of an inappropriate stellar brightness profile or the neglect of acceleration effects,
which include contributions due to the revolution of the Earth around the Sun or the orbital motion within the binary
lens or a possible binary source, leads to well-distinguishable characteristic systematics in the model residuals.
Although acceleration is unlikely to cause significant effects on the lightcurve for most
events, if used as a free model parameter, however, it blurs the measurement of the stellar brightness profile if the
duration of the caustic passage cannot be determined with sufficient accuracy, which is the case if
only the surroundings of the end of the caustic exit are sampled. Nevertheless, the related
degeneracy can be limited by applying a reasonable upper
limit to the absolute amount of acceleration resulting from an
assessment of the dynamical properties of the binary lens and the Earth's motion at the time of observation.

\section*{Acknowledgments}
This work has been made possible by postdoctoral support 
on the PPARC rolling grant
PPA/G/O/2001/00475.

\bibliographystyle{mn2e}
\bibliography{fld}

\end{document}